\documentstyle[aas2pp4]{article}

\def\kms{km s${}^{-1}$}

\def\a{$\alpha$}
\def\ab{$\sim$}
\def\yz{Yusef-Zadeh}
\def\p{$\pm$}
\def\gusten{G\"{u}sten}
\def\NT{Northern Thread}
\def\ST{Southern Thread}
\def\l{$\lambda$}

\begin{document}
\title{A Radio Polarimetric Study of the Galactic Center Threads}
\author{Cornelia C. Lang\altaffilmark{1,2}, Mark Morris\altaffilmark{2}, and Luis Echevarria\altaffilmark{3}}     
\altaffiltext{1}{National Radio Astronomy Observatory, Box 0, Socorro,
NM 87801; email: clang@nrao.edu}
\altaffiltext{2}{Division of Astronomy, 8371 Math Sciences Building, Box
951562, University of California at Los Angeles, LA, CA 90095-1562}
\altaffiltext{3}{Department of Physics \& Astronomy, Arizona State University, Tempe, AZ}
\begin{abstract}
The Very Large Array has been used to carry out a multifrequency, 
polarimetric 
study of the non-thermal filaments (NTF's), G0.08+0.15, and 
G359.96+0.09, also known as the Northern and Southern Threads (Morris \& \yz~1985).  
These linear structures have been observed at \l=20, 6, 3.6 and 2 cm,
with high enough spatial resolution to be resolved for the first time at \l6 and \l3.6 cm. 
The \l20 cm image reveals a wealth of new detail in the radio sources
lying within the inner 60 pc of the Galaxy. The Southern
Thread has a prominent split along its length, similar to splitting at
the ends of previously studied NTF's.  In addition to the
prominent Northern and Southern Threads, there are several elongated
features that resemble NTF's, but extend for only 5$-$7 pc. 
With resolutions as fine as 2\arcsec, the \l3.6 and \l6 cm images
reveal a high degree of continuity and little substructure internal to
the filament. However,  the width of the Northern Thread varies along its
length between \ab4\arcsec~and \ab12\arcsec~(0.15$-$0.5 pc),
and becomes markedly diffuse at its NW extremity. 
The spectral index of the \NT~has been determined over a broad range
of frequencies.  Its flux density falls with
frequency, ($\alpha$=$-$0.5) between \l90 and \l6 cm (where S$_{\nu}$
$\propto$ $\nu$$^{\alpha}$), and becomes much steeper (\a=$-$2.0)
between \l6 and \l2 cm.  The spectral index does not
vary significantly along the length of the \NT, which implies either
that the diffusion timescale for the emitting electrons is less than
their synchrotron lifetime, or that the emitting electrons are
reaccelerated continuously at multiple positions along the filament.
Because of the lack of spectral index variation, we have not located the source of relativistic electrons. 
Polarization observations at \l6 and \l3.6 cm confirm the non-thermal
nature of the emission from the Northern Thread. The fractional
polarization in the \NT~reaches 70\% in some regions, although
the polarized emission is patchy.  
Large rotation measures (RM~$\gtrsim$~2000 rad m$^{-2}$) have been observed
with irregular variations across the filament; typical values 
 of RM are 100$-$2000 rad m$^{-2}$. The lack of any apparent pattern in the 
distribution of RM suggests that the Faraday rotating medium is not 
physically assocated with the Northern Thread since the filament itself is so highly ordered. 
The intrinsic
magnetic field in the Northern Thread
is predominantly aligned along its long axis. 
The data on the Southern Thread were less conclusive:  the Southern
Thread was not detected in total intensity at \l6 or \l3.6 cm, while the polarized emission at \l3.6 cm arising from the Southern Thread is
evident, with a bifurcated structure similar to that detected
in total \l20 cm intensity.
\end{abstract}

\section{Introduction}
In the past fifteen years, a number of unusual non-thermal
filaments (hereafter, NTF's) 
have been observed with the Very Large Array
(VLA)\footnotemark\footnotetext{The VLA is a facility of the National
Science Foundation, operated under a cooperative agreement with the Associated
Universities, Inc.} in the inner regions of the Galaxy.
These peculiar NTF's share the following characteristics (see reviews by Morris 1996;
Morris \& Serabyn 1996): (1) unique to the
Galactic Center and found only within a projected distance of 150
pc from SgrA (assuming hereafter a distance of 8.0 kpc to the Galactic
center; Reid et al. 1993), (2) lengths of tens of pc, yet widths of
only $<$ 0.5 pc, (3) orientation
essentially perpendicular to the Galactic plane within \ab20\arcdeg, (4) strong linear
polarizaton (up to \ab50\%), (5) spectral indices negative,
flat or slowly rising,  (6) rotation measures often $>$ 1000 rad m$^{-2}$,
and (7) the intrinsic magnetic fields are
aligned along the long axis of the filament. In addition, recent observations have
found that each
sufficiently well-studied NTF appears to be associated with the
ionized edge of a molecular cloud.

The synchrotron nature of the radio emission from the NTF's was first
proposed by Yusef-Zadeh, Morris, \& Chance (1984), who discovered the
NTF's in the Galactic center Radio Arc at $\ell$=0\fdg2.  Subsequent polarization studies
of Tsuboi et al. (1986, 1995), Yusef-Zadeh
\& Morris (1987b), and Reich (1994) 
confirmed that the emission was strongly linearly polarized, and that
the internal magnetic field orientations are parallel to the Radio Arc
NTF's. The more isolated Northern and Southern Threads (G0.08+0.15
and G359.96+0.09)  have been observed previously at \l=6, 20, and 90 cm with
fairly low resolution (Morris \& Yusef-Zadeh 1985; Anantharamaiah et
al. 1991). These studies show that the 
spectral index of the \NT~between \l90 and \l20 cm is negative, \a~=~$-$0.6\p0.1, 
and that the spectral index of the \ST~is nearly flat:
\a~=~$-$0.04\p0.03. Recent high resolution and polarimetric studies of 
the other NTF's, G359.1$-$0.2, the ``Snake'' (Gray et al. 1995) and G359.54+0.18 (Yusef-Zadeh et al. 1997), reveal fine
sub-filamentation along the lengths of these NTF's and
polarization properties similar to that of the Radio Arc
NTF's. Observations at \l20 cm
have also revealed a bifurcated NTF associated with the SgrC H II
complex (Lizst \& Spiker 1995). 

There are a number of unresolved questions raised by the unusual
Galactic center NTF's.  The most puzzling and fundamental of these is,
what is the source of relativistic
particles and the mechanism for their acceleration? The apparent associations between NTF's and both
ionized and molecular gas provide a potential clue:  is this association
somehow related to, and necessary for, the generation of NTF's?
Finally, what can the NTF's suggest about the overall magnetic field
configuration at the Galactic center? and what can confine such narrow structures?

Several models of NTF generation have been proposed (see Morris
1996) and are summarized here. 
Benford (1988) and Morris \& Yusef-Zadeh (1989) point
out that the v$\times$B electric field induced at the surface of a
molecular cloud moving at high velocity through a uniform magnetic
field may be 
sufficient to accelerate electrons to relativistic energies.  Heyvaerts et al. (1988) propose that the NTF's result
from magnetic reconnection occuring at the interface of a magnetic
loop expanding from the Galactic nucleus with a magnetic wall
surrounding the central regions.  Magnetic reconnection at the
interface between
adjacent magnetic pinches has also been invoked to explain electron acceleration in the bundle of NTF's which compose the
Radio Arc (Lesch \& Reich 1992).  A recent model proposed by
Rosner \& Bodo (1996) suggests that the NTF's are formed as a result of
the interaction of a fast, ionized stellar wind (emanating from a star
with a high mass loss rate) with the ambient magnetic
field. They argue that electrons could be accelerated to relativistic
energies across the magnetopause, and loaded onto magnetic flux tubes, thus
illuminating the NTF's. 
A more exotic idea proposed by Chudnovsky et al. (1986) is that the
NTF's are manifestations of superconducting
cosmic strings interacting with the magnetized plasma near the
Galactic center. 

Recent work on the NTF's has been aimed at determining the nature and
the ubiquity of the interactions between the NTF's and the associated ionized and
molecular gas.  Based on high resolution CS(J=2$-$1) observations,  
Serabyn and Morris (1994; see also Serabyn \& \gusten~1991) propose that the source of the electrons and
their acceleration in the Radio Arc occurs at the intersection of
these NTF's with the Sickle H II region and the underlying molecular
cloud.  Their observations showed that the molecular gas is distributed
in clumps at locations where both the NTF's and ionized gas are
present, and where changes in both brightness and continuity of the Radio
Arc NTF's occur (Yusef-Zadeh \& Morris 1987c). Serabyn \& Morris
(1994) propose that at these positions, reconnection occurs
between the magnetic field internal to the molecular cloud and the
strong, external field, and may accelerate the electrons to
relativistic energies, thus launching them to diffuse along the
external field lines.  These field lines are then illuminated by
synchrotron radiation and thereby form an NTF. 
Morphological connections between the radio continuum,
recombination line and molecular line observations in many NTF systems indicate
that the model proposed by Serabyn \& Morris (1994) may be widely
applicable amongst the NTF's (e.g. Lizst \& Spiker 1995, Uchida et al. 1996,
Staguhn et al. 1998). 

Here we present high resolution, multi-frequency data on the first isolated NTF's discovered, the Northern
and Southern Threads, in order to test whether they share the
properties as the other NTF's. Radio polarimetry has also been
done in order to elucidate the synchrotron nature of these features,
to determine their
intrinsic magnetic field orientation, and to characterize the
foreground Faraday screen.

\section{Observations \& Results}
The multi-frequency observations presented here originate from several
different epochs and configurations of the VLA. The frequency, phase
center, and array information for all observations is
summarized in Table 1.  Standard AIPS procedures for calibration,
editing, and imaging were used for all data. 3C 286 was used in all
instances for flux calibration, and NRAO 530 (1733$-$130) and 1748$-$253
were used for phase and polarization calibration.

\begin{deluxetable}{cccccc}
\singlespace
\tablecaption{Observed Fields\label{tab:obs}}
\tablehead{
\colhead{Field} &
\colhead{Array} &
\colhead{Date} &
\colhead{Frequency (GHz)}&
\multicolumn{2}{c}{Phase center}\\
\cline{5-6}
\colhead{Designation} & \colhead{} & && \colhead{$\alpha$ (J2000)}& \colhead{$\delta$ (J2000)}}
\tablewidth{0pt}
\tablecolumns{7}
\startdata
Sgr A Complex&CnB &27 June 1985&1.446 & 17 44 40.5&$-$28 51 14.0\\
& CnB & 6 Oct 1986 &''&'' &''\\
& CnB & 20 May 1989 &''&'' &''\\
& DnC & 30 May 1987 &'' &'' &''\\
& BnA & 12 Feb 1987 &'' &'' &'' \\
Northern Thread 6 & CnB & 20 May 1989& 4.585, 4.885&17 45 12.4&$-$28 46 46.6\\
& D& 30 May 1987&4.585, 4.885&'' &''\\
&  BnA&28 Sept 1995&''&''\\
&  DnC&22 Jan 1995&'' &''\\

Northern Thread 3.6(a)&BnA&28 Sept 1995&8.085, 8.465&17 45 12.4&$-$28 46 46.6\\
& DnC & 22 Jan 1995&''&''&'' \\ 
Northern Thread 3.6(b)&BnA &28 Sept 1995&8.085, 8.465&17 43 30.4
&$-$28 50 03.9\\
 &DnC & 22 Jan 1995&'' &''&''\\
Northern Thread 2(a)& D &30 May 1987& 14.460 &17 45 12.3&$-$28 48 46.6\\
Northern Thread 2(b) & D &30 May 1987&''& 17 45 19.1&$-28$ 48 26.1\\
Northern Thread 2(c) & D &30 May 1987&''& 17 45 27.9& $-$28 49 40.0\\
Southern Thread 6&BnA& 28 Sept 1995&4.585, 4.885&17 45 09.8& $-$28 55 07.8\\
& DnC & 22 Jan 1995&'' &'' &''\\
Southern Thread 3.6&BnA & 28 Sept 1995 &8.085, 8.465&17 45 09.8& $-$28 55 07.8\\
 &DnC&22 Jan 1995&''&''&''\\
\enddata

\end{deluxetable}
\begin{deluxetable}{lccl}
\singlespace
\tablecaption{Census of Linear Features Perpendicular to the Galactic Plane Detected in Figure 2}
\tablehead{
\colhead{Galactic} & \multicolumn{2}{c}{Equatorial Coordinates} &
\colhead{notes}\\
\cline{2-3}\\
\colhead{Coordinates} & \colhead{$\alpha$ (J2000)}& \colhead{$\delta$ (J2000)} &
\colhead{}}

\tablewidth{0pt}
\tablecolumns{4}
\startdata
G0.16-0.15&17 46 00&$-$28 47 00& Radio Arc\\
G0.15-0.07& 17 46 15& $-$28 51 30 & 'steep spectrum' filament S of Radio Arc\\
G0.08+0.15&17 45 15 &$-$28 48 00 & Northern Thread\\
G0.08+0.02&17 45 45& $-$28 51 30& extension of the Northern Thread\\
G359.96+0.09&17 46 05 & $-$28 55 00 & Southern Thread\\
G359.79+0.17&17 44 30 &$-$29 02 00 & curved filament\\
G359.98-0.11& 17 46 00 &$-$29 00 40 & streak\\
G359.88-0.07& 17 45 35 & $-$20 04 00 & streak\\
G0.02+0.04&17 45 30 & $-$28 53 40 & streak\\
G0.06+0.06&17 45 35 & $-$28 51 40& quasi-linear features\\

\enddata
\end{deluxetable}

\subsection{\l20 cm Total Intensity Image}

Figure 1 is a schematic diagram of the sources located in the inner 50
pc of the Galaxy, shown in Figure 2 and
discussed throughout this paper.  Figure 2 is the \l20 cm continuum
image with a resolution of \ab6\arcsec, made using the
maximum entropy method of deconvolution, VTESS, in AIPS.

\subsubsection{The Non-Thermal Filaments}
Both the Northern and Southern Threads are prominent features in
Figure 2.  The Northern Thread crosses the Arched Filaments (W1 and
W2), and extends 12\arcmin~(30 pc) to the W from that point and 3\arcmin~(8 pc) to the
E, with a total length of \ab40 pc.
At the E extremity of the Northern Thread, (\a,$\delta$)$_{J2000}$=17 45
45, $-$28 51 30, there is another parallel filament of length
\ab4\arcmin~(10 pc), but it is displaced 20\arcsec~to the S of the
Northern Thread. It appears as if this second filament is a
continuation of the Northern Thread.  At \l20 cm, the \NT~exhibits very little variation in brightness or
 width and shows continuity along its length. It has,
 however, a gentle curvature, from where it is essentially
 parallel to the Radio Arc, to its W extremity, where it is oriented
 NW with a change of about 20\arcdeg~from its original orientation.
 At the W extremity, the \NT~appears bifurcated: a second faint
filament runs parallel for \ab2\arcmin~(5 pc), and also becomes diffuse to the W.

The \l20 cm map provides new detail in the Southern
Thread. This NTF extends for
\ab11\arcmin~(28 pc), but is indistinct toward its E extremity, in the halo of
the Sgr A complex.   
The Southern Thread has an average width of \ab10\arcsec~(0.4 pc) along its length,
and shows a clear split toward its center,  extending for
\ab4\arcmin~(10 pc) along its length. 
The two parallel strands into which the Southern Thread is
divided have about the same brightness, giving the impression of two
filaments. Figure 3 shows a detail of the Southern Thread from the
\l20 cm image shown in Figure 2. Although it is very faint, the more
northern filament of the Southern Thread does
appear to continue to at least the edge of the diffuse halo of the Sgr
A complex.  In addition, a possible third parallel filament is present
just S of the split.

The NTF's of the Radio Arc also stand out in this image and show the
sub-filamentation and fine structure which has been described by
Yusef-Zadeh \& Morris (1987b). However, in this image, the Radio Arc is at the edge of the
primary beam here (30\arcmin~at \l20 cm), and the NTF's are therefore not as
well characterized as in previous images.
Another feature in the \l20 cm image which resembles the NTF's is
G359.79+0.17, located W of the Sgr A complex (Yusef-Zadeh 1986;
Yusef-Zadeh \& Morris 1987a; Anantharamaiah et al. 1991; Morris 1996);
a detail of
this filament is shown in Figure 4.  At the diffuse, eastern edge of this filament, there appear to
be multiple, parallel filaments, as described above for both the Northern and Southern
Threads.  This substructure has been confirmed by higher resolution,
polarimetric observations, which also demonstrate that G359.79+0.17
shares all the characteristics of the NTF's (Lang \& Anantharamaiah, in
prep). 
\clearpage

\vbox{%
\begin{center}
\leavevmode
\hbox{%
\epsfxsize=15.0cm
\epsffile{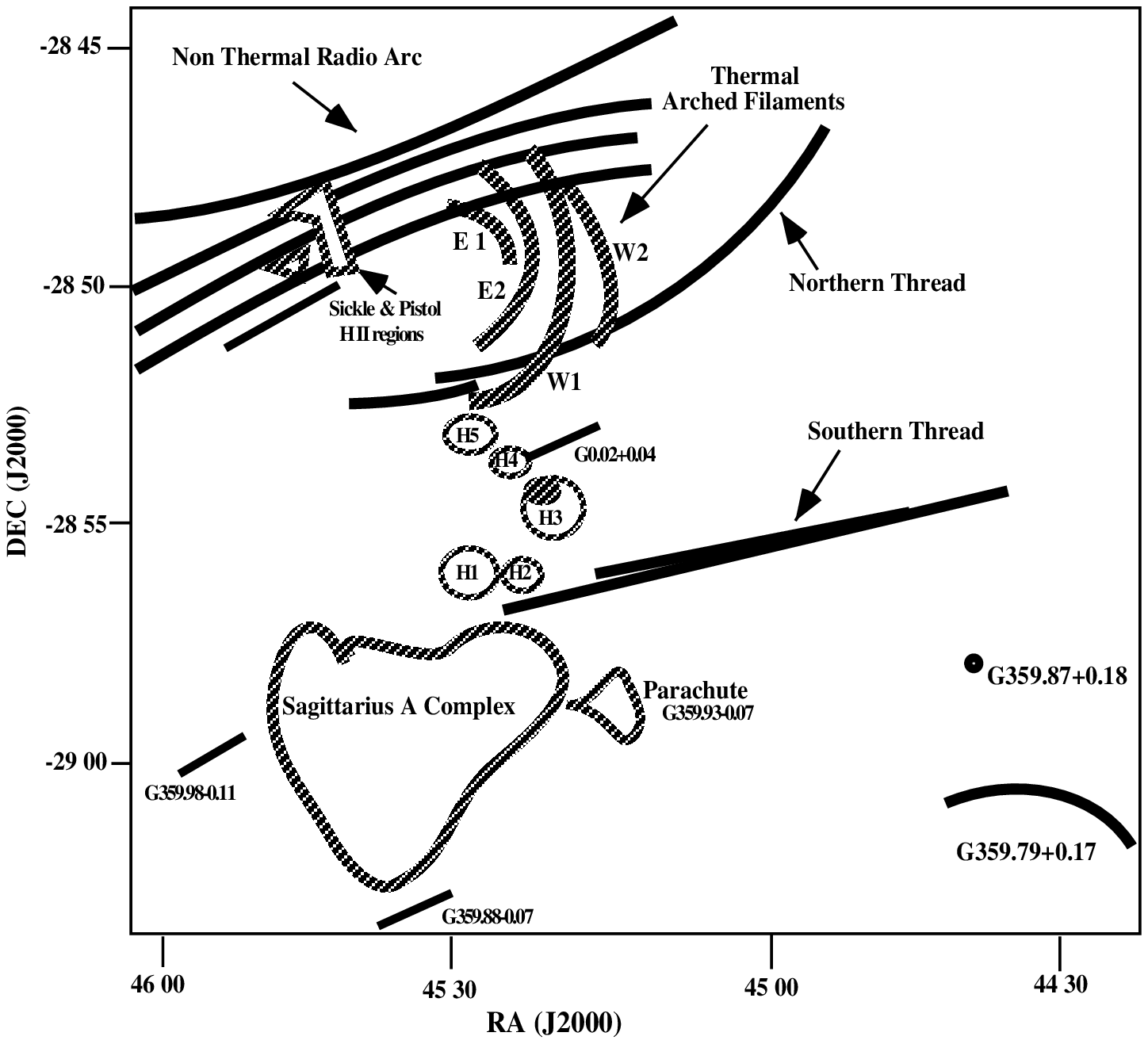}}
\begin{small}
\figcaption{\small Schematic representation of the sources that are
present in a 60 pc region of the Galactic center shown Figure 2 and discussed throughout this paper.}
\end{small}
\end{center}}

\clearpage

\vbox{%
\begin{center}
\leavevmode
\hbox{%
\epsfxsize=15.0cm
\epsffile{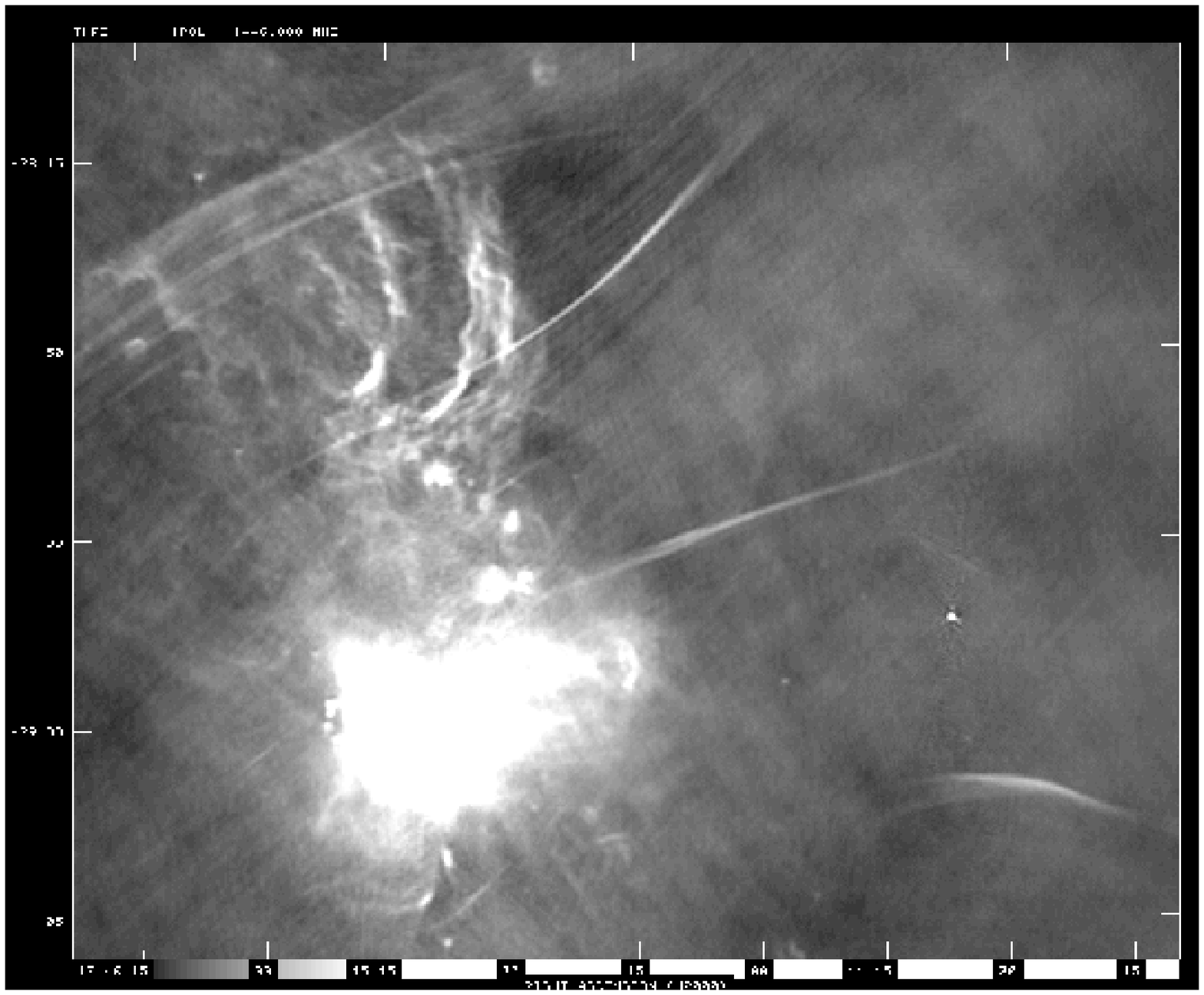}}
\begin{small}
\figcaption{\small \l20 cm continuum image of a 60 pc region of the Galactic
center, mostly at postive Galactic latitudes and longitudes.   The resolution
of this image is 5\farcs90 $\times$ 5\farcs5, PA=80\arcdeg. The image
was made with uniform weighing and has been corrected for the primary
beam attenuation. The rms noise level in this image is 0.5 mJy beam$^{-1}$.}
\end{small}
\end{center}}

\clearpage

\vbox{%
\begin{center}
\leavevmode
\hbox{%
\epsfxsize=7.5cm
\epsffile{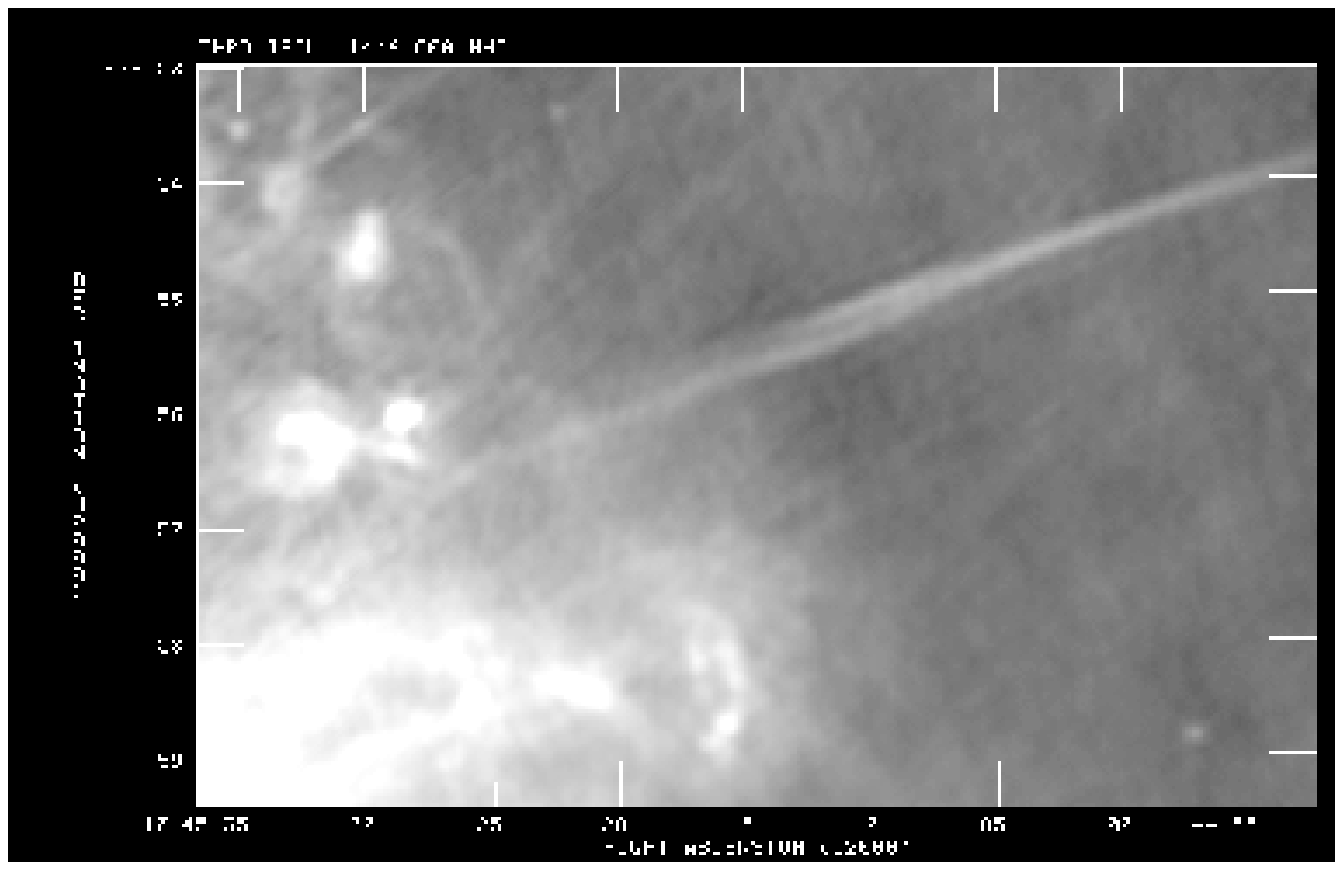}}
\begin{small}
\figcaption{\small Detail of the Parachute, Southern Thread, and H sources (see Figure 1 for locations) from the \l20 cm image shown in Figure 2.}
\end{small}
\end{center}}

\vbox{%
\begin{center}
\leavevmode
\hbox{%
\epsfxsize=7.5cm
\epsffile{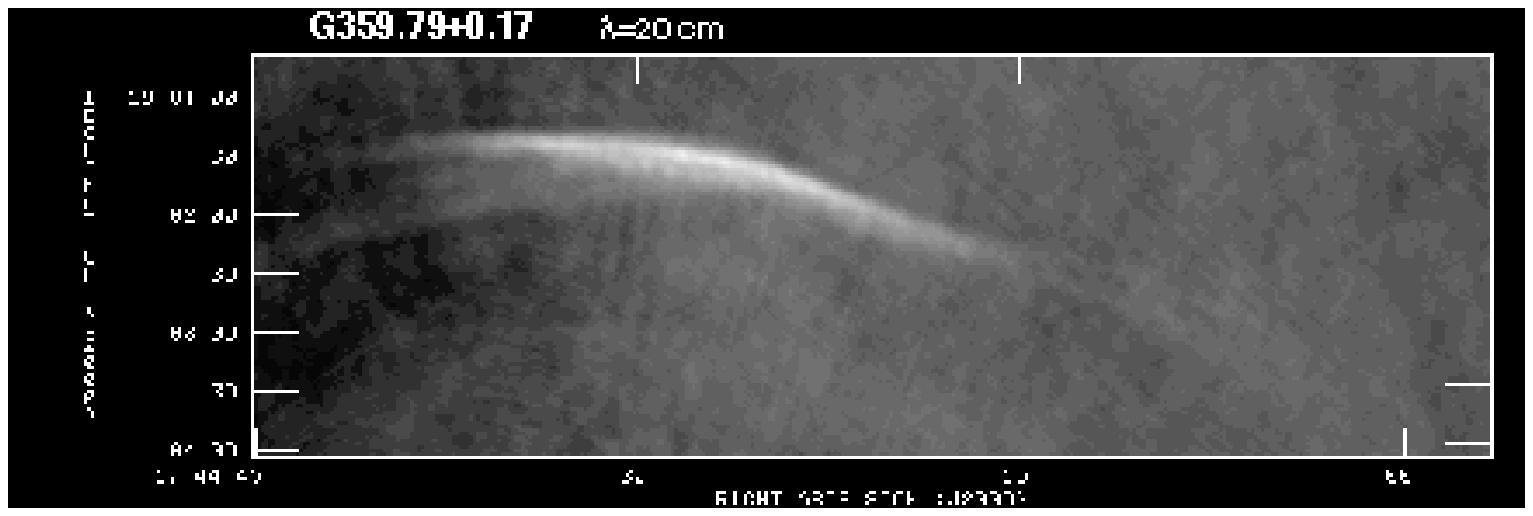}}
\begin{small}
\figcaption{\small A detail from Figure 2 showing the filamentary source G359.79+0.17.}
\end{small}
\end{center}}

\subsubsection{Additional Linear Structures: Streaks}
Although the Threads are the most striking linear features in Figure
2, there are a number of other linear structures
present which resemble the NTF's, although they only extend for
2$-$3\arcmin~(5$-$7.5 pc). We refer to these features as ``streaks'',
following Yusef-Zadeh and Morris (1987a) who first identified several
of them.   A census of all
filament-like sources detected in Figure 2, including streaks and 
known NTF's, is given in Table 2.
Similar to the NTF's, the streaks are elongated perpendicular to the Galactic
Plane and have comparable surface brightnesses. However, polarimetric
observations are crucial for determining whether or not these streaks are
the same class of objects as the NTF's. We note the following streaks
in Figure 2 which are labelled in Figure 1:
(1) G359.98-0.11, E of the SgrA complex; it is conceivable that this feature is a continuation of the
NTF; (2) G359.88-0.07, S of the SgrA complex; (3) G0.02+0.04, S of the
Arched Filaments, which appears to be associated with the H II
region H4.  Contour images of two of these streaks are shown in
Figures 5 and 6.  
\vbox{%
\begin{center}
\leavevmode
\hbox{%
\epsfxsize=7.5cm
\epsffile{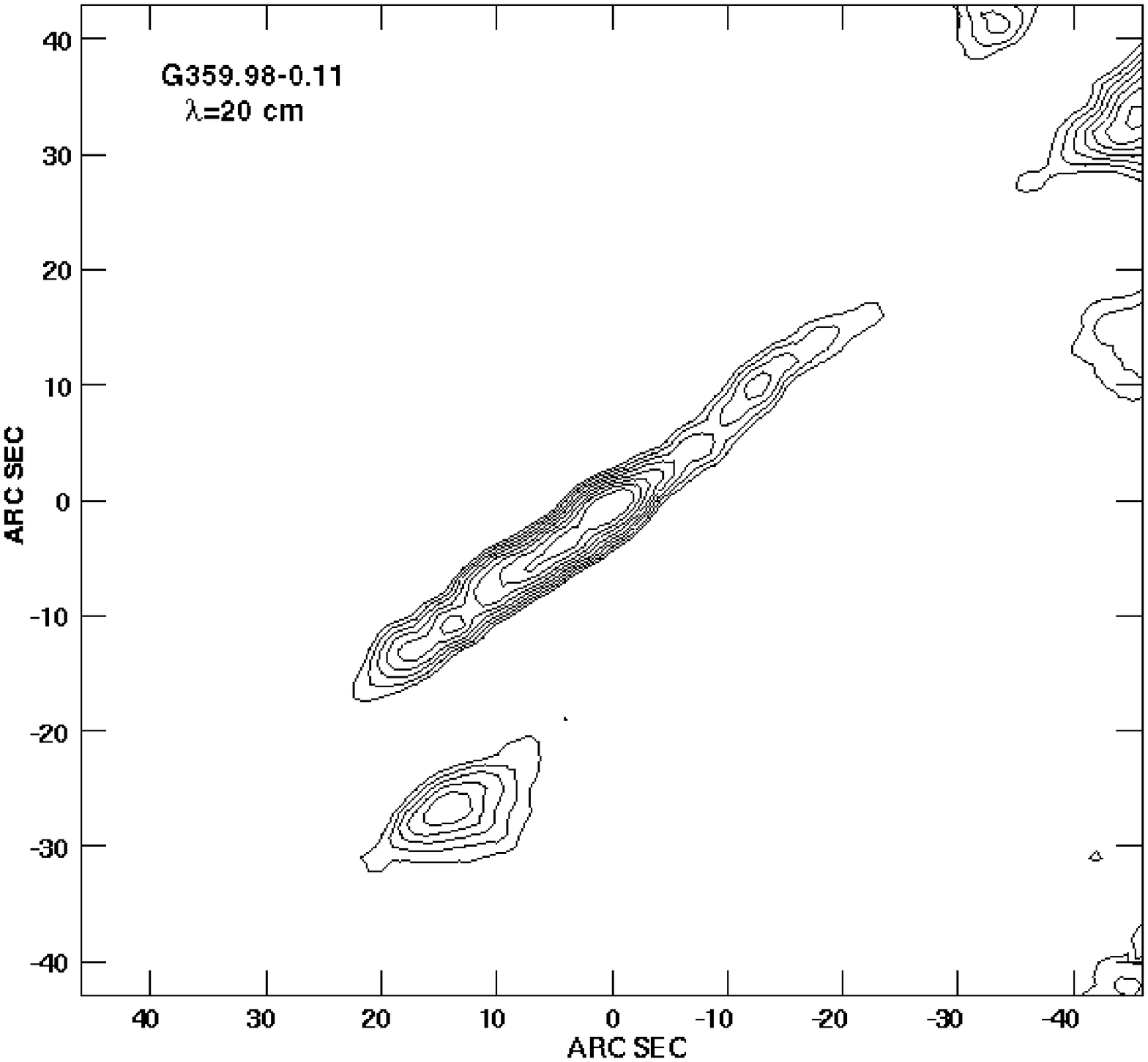}}
\begin{small}
\figcaption{\small \l20 cm contour image of the elongated structure G359.98-0.11, classified as a 
``streak''. The contours in this image represent 2, 3, 4, 4.5, 5, 5.5,
6, 6.5, 7, 8, 9, 10, 11, 12, 13, 14 mJy beam$^{-1}$ levels, and the
image is centered at (\a, $\delta$)$_ {2000}$=17 42 25, $-$29 03 28.}
\end{small}
\end{center}}

\vbox{%
\begin{center}
\leavevmode
\hbox{%
\epsfxsize=7.5cm
\epsffile{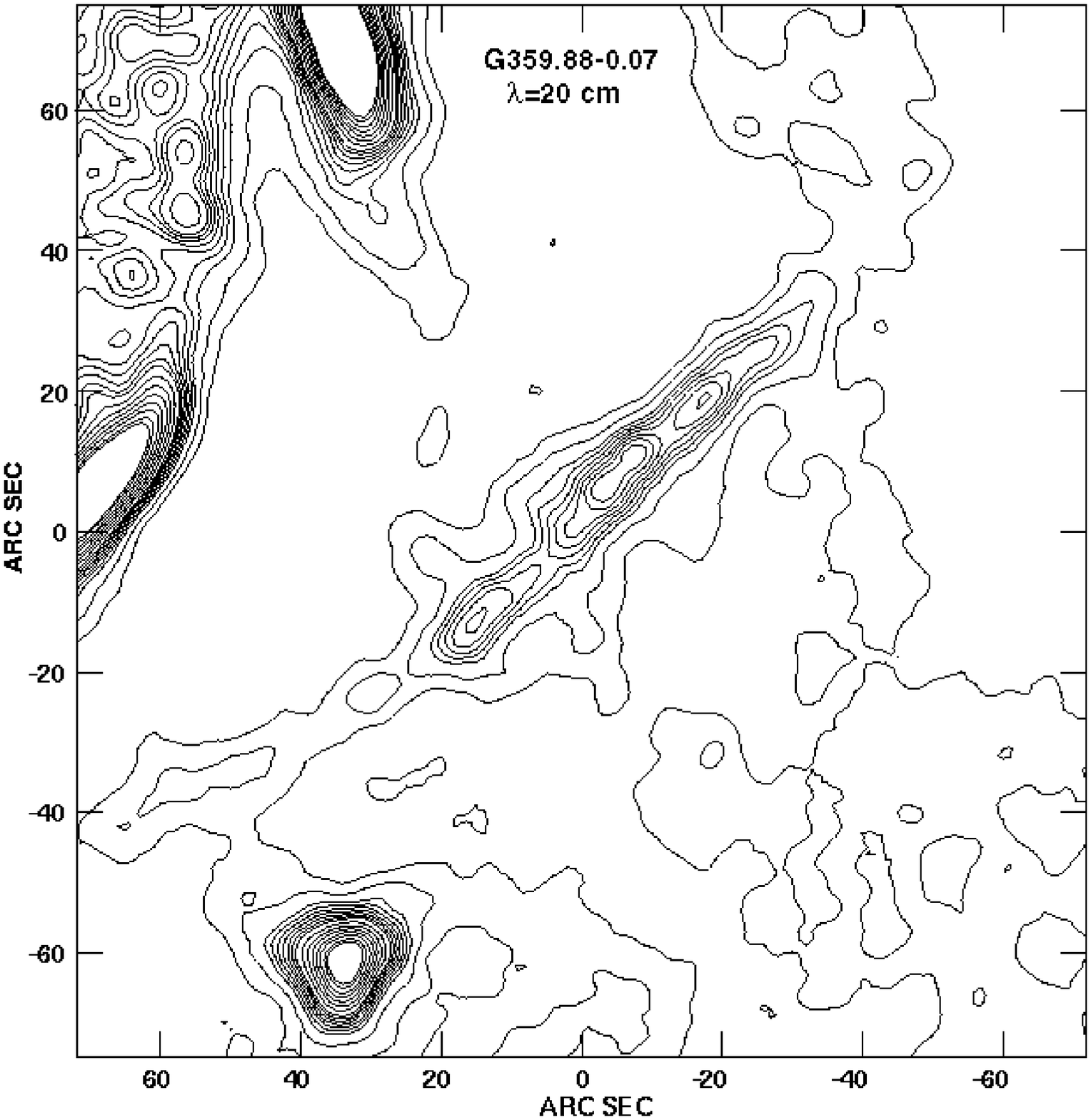}}
\begin{small}
\figcaption{\small \l20 cm contour image of the streak,
G359.88-0.07. The contours in this image represent 2.2, 2.4, 2.6, 3.0,
3.2, 3.4, 3.6 mJy beam$^{-1}$, and the image is centered at (\a, $\delta$)$_{J2000}$=17 42 50, $-$28 59 31.}
\end{small}
\end{center}}

\begin{deluxetable}{lccl}
\singlespace
\tablecaption{Sources detected near the Galactic center in these data}
\tablehead{
\colhead{Galactic} & \multicolumn{2}{c}{Equatorial Coordinates} &
\colhead{notes}\\
\cline{2-3}\\
\colhead{Coordinates} & \colhead{$\alpha$ (J2000)}& \colhead{$\delta$ (J2000)} &
\colhead{}}
 
\tablewidth{0pt}
\tablecolumns{4}
\startdata
G359.93-0.07& 17 45 20 &$-$28 58 20&''Parachute''; thermal source\\
G359.87+0.18 &17 44 37 &$-$28 57 09& FRII Galaxy (Lazio et al. 1999)\\
G0.09+0.17& 17 45 08 &$-$28 46 17& S$_{3.6}$=1.0$\pm$0.4 mJy;
S$_{6}$=2.0$\pm$0.6 mJy\\ 
 
\enddata
\end{deluxetable}
 
In addition to the streaks, a series of quasi-linear structures are
apparent in Figure 2 within 1\arcmin~of (\a, $\delta$)$_{J2000}$=17 45 35,
$-$28 51 40.  These structures, first noted by Morris \& \yz~(1989),
are aligned perpendicular to the Galactic plane.  Like the nearby
Arched Filaments,
these quasi-linear features may be thermally emitting structures,
although the similarity of their orientation to the that of the nearby
NTF's and streaks suggest that they have also been shaped by the
magnetic field.
Near the Radio Arc, we note an
additional filament,  previously reported by Anantharamaiah et
al. (1991) and known as the 'steep spectrum filament'. This filament
extends for 3\farcm5, running exactly parallel to and S of the
Radio Arc NTF's. It may be associated with the
Radio Arc, but as Anantharamaiah et al. (1991) point out, its spectral
index is negative, \a=$-$0.4, compared to that of the Radio Arc:
\a=+0.3. 

\subsubsection{Other Sources}
The group of compact sources located between SgrA
and the Arched Filaments are labelled in Figure 1 as
H1$-$H5, after Yusef-Zadeh \&~Morris (1987a) and are known
Galactic center H II regions. Several of these sources are highlighted
in Figure 3. In particular, the H II region H3 shows a continuous ring-like structure. Radio recombination line studies reveal
that these H II regions are likely associated with the $-$30 \kms~molecular cloud
(Serabyn \& \gusten~1987), and can be characterized by T$_e$, n$_e$,
and $\Delta$v$_{FWHM}$, typical for Galactic center H II regions (Zhao
et al. 1993). 
The source located at the W edge of the Sgr A complex (G359.93-0.07,
or the ``Parachute''), also evident in Figure 3 may be related to the
``Streamers'' which connect it morphologically to SgrA (Yusef-Zadeh \& Morris 1987a).
Finally, the point source G359.87+0.18, detected earlier at \l20 and \l90 cm
(\yz~\&~Morris 1987a; Anantharamaiah et al. 1991) is
also apparent in Figure 2 at (\a, $\delta$)$_{J2000}$=17 44 37.2, $-$28 57 08. Lazio et al. (1999) have classified this
source as a Faranoff-Riley II radio galaxy, based on VLA observations
in the range of 0.33 to 15 GHz and an HI absorption spectrum.   
Table 3 summarizes the sources discussed in this section. 

\subsection{Total Intensity Images at \l6, \l3.6 cm \& \l2 cm}
Figure 7 shows the high resolution (\ab2\arcsec) $\lambda$6 cm image of the Northern
Thread and the Arched Filaments. 
At \l6 cm, the Northern Thread extends \ab10\arcmin~(25 pc), from
its E extreme where it intersects the thermal Arched Filaments,  to its W
extreme where it becomes diffuse.  We note a point source
located near the W end of the Northern Thread (see Table 3).  
In previous observations (Morris \& \yz~1985; Anantharamaiah et al. 1991), the Northern Thread appears
wider, due to lower resolution.  However at \l6 cm, the \NT~is
fully resolved for the first time. It is obvious in Figure 7
that the width of the Northern Thread changes along its length. Just E of the position
where the \NT~crosses the Arched Filaments, it is narrowest, $<$
4\arcsec~(0.15 pc).  It becomes broader toward the W, with the largest
width near its W extreme, \ab12\arcsec~(0.5 pc).  
The bifurcation of the W end of the \NT~detected at \l20 cm in Figure
2 is not observed at \l6 cm or
\l3.6 cm.  In addition, no apparent kinks or bendings occur along the
\NT, and there is no strong evidence for the type of obvious
sub-filamentation that has been observed in the Snake NTF
(Gray et al. 1995), G359.54+0.18 (\yz~et al. 1997), and
in the splitting of the Southern Thread. 

\vbox{%
\begin{center}
\leavevmode
\hbox{%
\epsfxsize=7.5cm
\epsffile{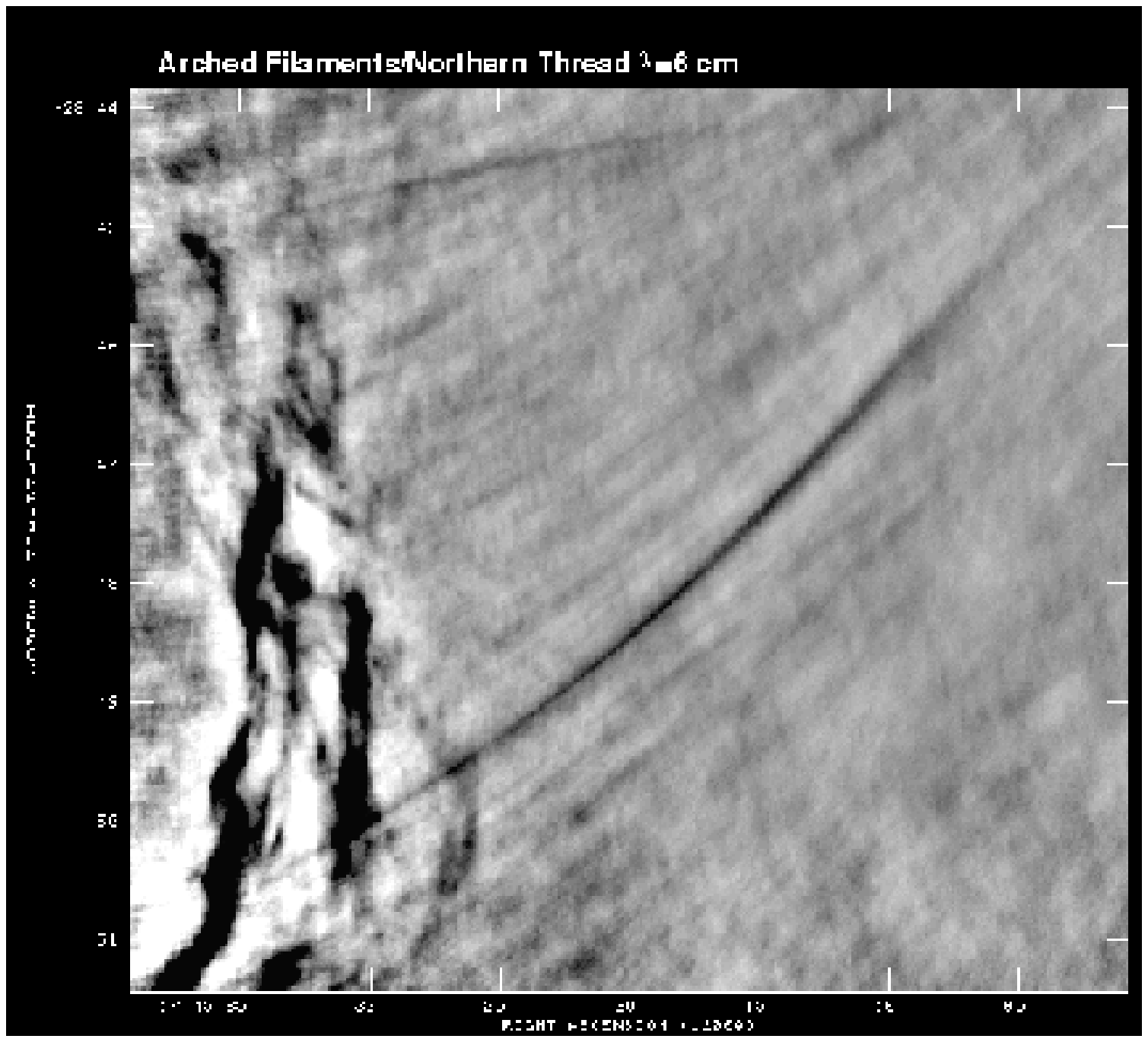}}
\begin{small}
\figcaption{\small High resolution \l6 cm image shown in negative
greyscale of the Northern Thread and a portion of the Arched Filaments
H II regions. The image was made with uniform weighting and has a resolution of
2\farcs12 $\times$ 1\farcs73, PA=0\fdg91, and has been corrected for
primary beam attenuation.  The rms noise in this image is 0.1 mJy
beam$^{-1}$.}
\end{small}
\end{center}}

Figure 8 shows a mosaic of the \NT~at \l2 cm, made from three
overlapping fields. At such a high frequency, the Northern Thread can
be followed for \ab6\arcmin~(15 pc) toward the W of the Arched
Filaments. Total intensity images of the Southern Thread were also
made at \l6 and \l3.6 cm, and in both cases, the Southern Thread does
not stand out against the background of the SgrA complex. Upper limits
for total \l6 and \l3.6 cm intensity are 0.7 mJy beam$^{-1}$ and 0.8
mJy beam$^{-1}$ respectively. 

\vbox{%
\begin{center}
\leavevmode
\hbox{%
\epsfxsize=7.5cm
\epsffile{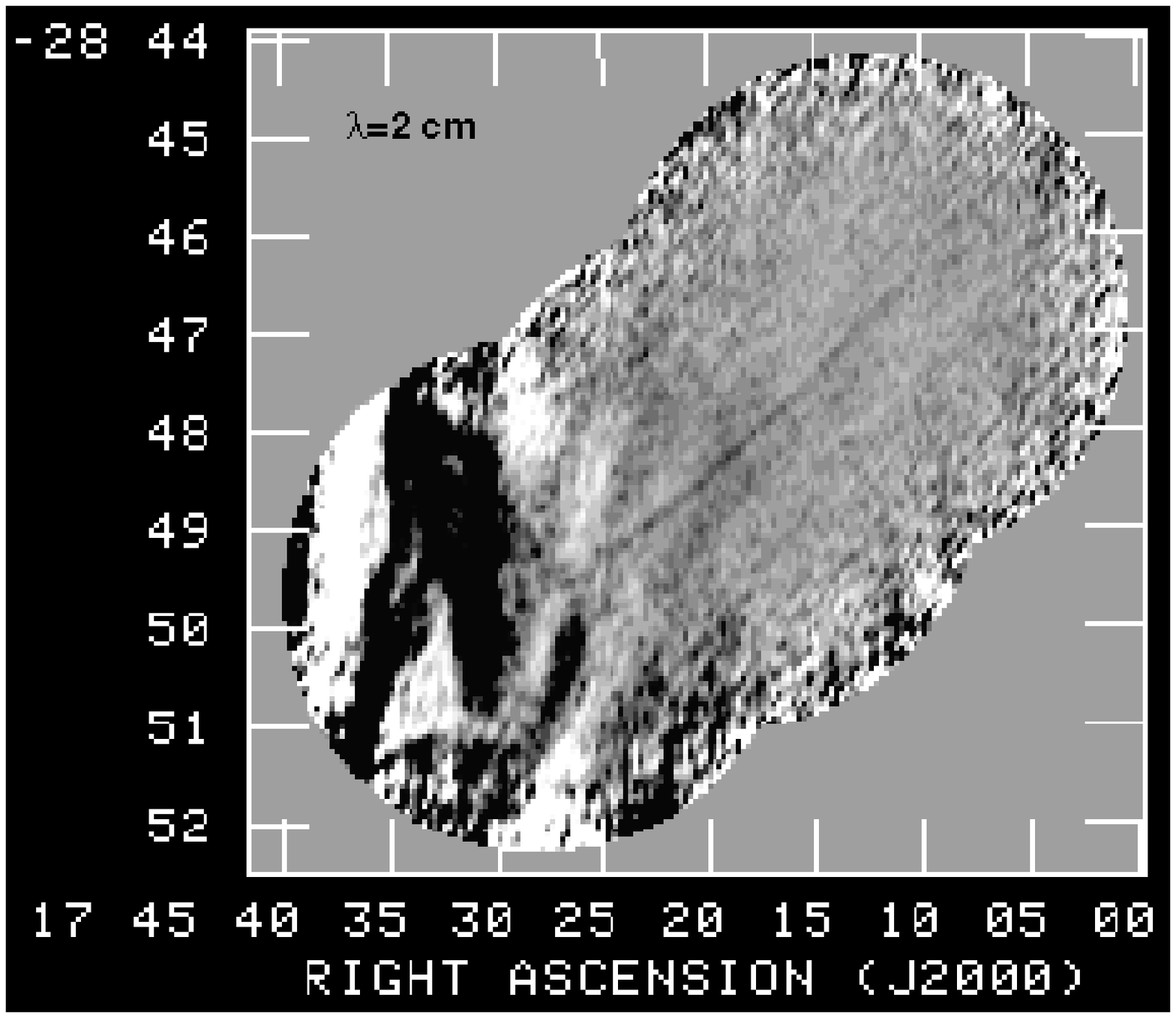}}
\begin{small}
\figcaption{\small \l=2 cm image of the Northern Thread and Arched Filaments
shown in negative greyscale.  This image is was mosaicked together
from three fields, and has a resolution of 4\farcs3 $\times$ 4\farcs3.}
\end{small}
\end{center}}

\subsubsection{Spectral Index of the Northern Thread}
The surface brightness of the ridge of the Northern Thread at each of
the four observed wavelengths is shown as a function of length along
the NTF in Figure 9.  From this figure, it is obvious that the surface
brightness decreases with shorter wavelengths, and that \l3.6 and \l2
cm data do not cover the entire extent of the Northern Thread.
Figure 10 shows the spectrum of the \NT,
sampled at different positions along its length. At 12 locations, the
intensity was averaged over a narrow (8\arcsec) region extending for
40\arcsec~perpendicular to the filament, in order to determine the
local zero-level. After this baseline was subtracted, the flux density
of the Northern Thread was measured at each position. The \l90 cm data were supplied by
Anantharamaiah (priv. communication). Between \l90 and \l6 cm,
\a\ab$-$0.5, and then a drastic steepening occurs between \l6 cm to
\l2 cm, where \a\ab$-$2.0.

Caution must be
exercised in comparing surface brightness between the different frequencies,
 since the the interferometer short spacings are less well sampled at
the higher frequencies than at the lower frequencies, and we could be
missing flux on the largest scales at \l3.6 and \l2 cm.  However, our
averaging procedure accounts for the different beam sizes and there is not strong evidence that significant flux has
been resolved out at higher frequencies, especially since the Arched
Filaments, which have a larger angular extent, are clearly detected at
all frequencies.

\vbox{%
\begin{center}
\leavevmode
\hbox{%
\epsfxsize=7.5cm
\epsffile{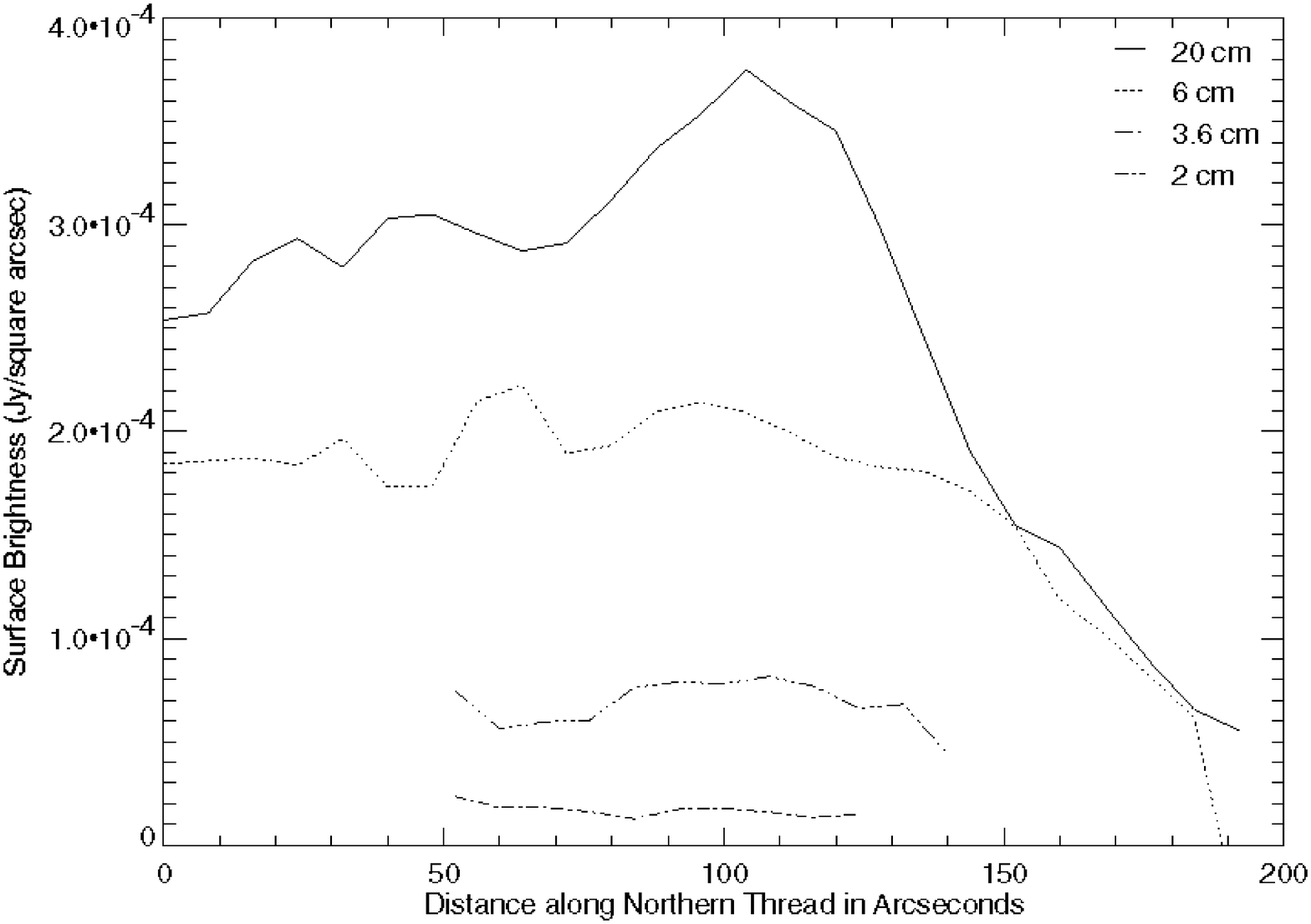}}
\begin{small}
\figcaption{\small Surface brightness of the Northern Thread for each of the four observed frequen
cies
as a function of length along the filament. The reference
position for \l6 and \l20 cm is
at (\a, $\delta$)$_{J2000}$=17 45 20, $-$28 50 00, and increasing
arcseconds westward from this position. At each frequency, the images have been adjusted
to a common resolution (6\farcs2 $\times$ 5\farcs5), and the units are Jy arcsecond$^{-2
}$.}
\end{small}
\end{center}}

\vbox{%
\begin{center}
\leavevmode
\hbox{%
\epsfxsize=7.5cm
\epsffile{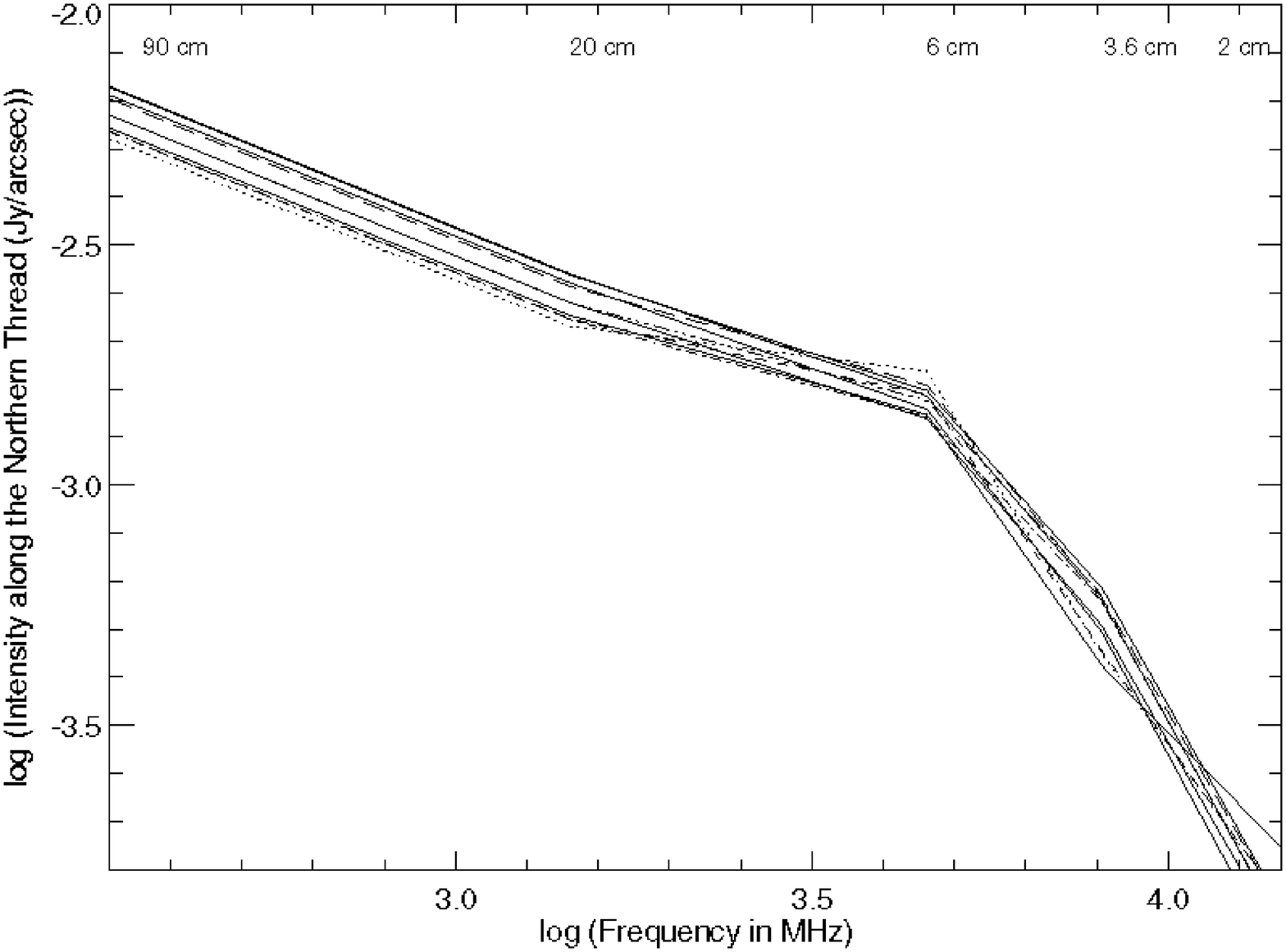}}
\begin{small}
\figcaption{\small The spectrum of the Northern Thread: log-log plot of intensity per arcsecond al
ong the Northern
Thread as a function of frequency.  The different linestyles
correspond to different positions along the NTF at which the intensity
was sampled. The sampling interval was 8\arcsec, starting along the
middle of the Northern Thread, and extending W for \ab180\arcsec.}
\end{small}
\end{center}}

 \vbox{%
\begin{center}
\leavevmode
\hbox{%
\epsfxsize=7.5cm
\epsffile{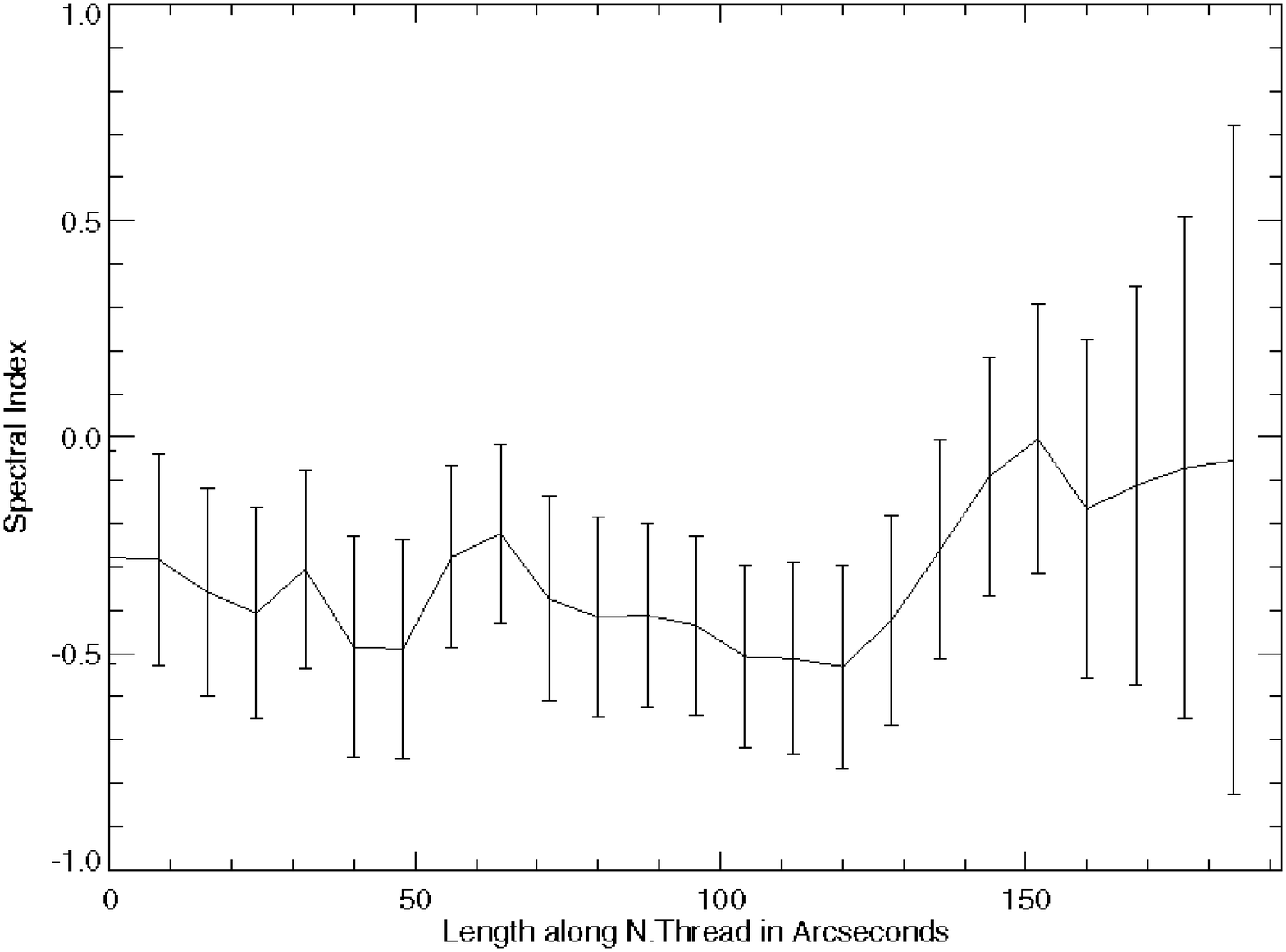}}
\begin{small}
\figcaption{\small \l20/6 cm spectral index as a function of length along the
Northern Thread, which is measured in arcseconds extending W
from the position (\a, $\delta$)$_{J2000}$=17 45 25, $-$28 50 00.}
\end{small}
\end{center}}
  
Figure 11 shows the \l20/6 cm spectral index as a function of length
along the Northern Thread. Since the smaller primary beams of the \l2 and
\l3.6 cm images do not include the entire extent of
the filament (see Figure 9), a four-wavelength determination of the spectral index
variation is not possible. Therefore, we determine the spectral index
between \l20 and \l6 cm
over an extent of 180\arcsec, starting at (\a, $\delta$)$_{J2000}$=17 45 25,
$-$28 50 00, and look for variations westward from this
position. Within the error bars, there is no significant variation of the spectral index along the length of the
Northern Thread.
Toward its diffuse W extent (between 100$-$180\arcsec), we note a possible flattening in the spectral index.

\section{Polarimetry Results}
\subsection{Northern Thread}

Stokes' Q and U images were made at \l6 cm (4.585 \& 4.885 GHz) and
\l3.6 cm (8.085 \& 8.465 GHz), and images of the polarization angle
(PA=$\case{1}{2}$Arctan($\case{U}{Q}$)) and polarized intensity (I$_p$=$\sqrt{Q^2 +
U^2}$) were created.  
The observed polarization angles at \l6 and \l3.6 cm of the Northern Thread are shown in Figures 12 and 13.
\clearpage

\vbox{%
\begin{center}
\leavevmode
\hbox{%
\epsfxsize=7.5cm
\epsffile{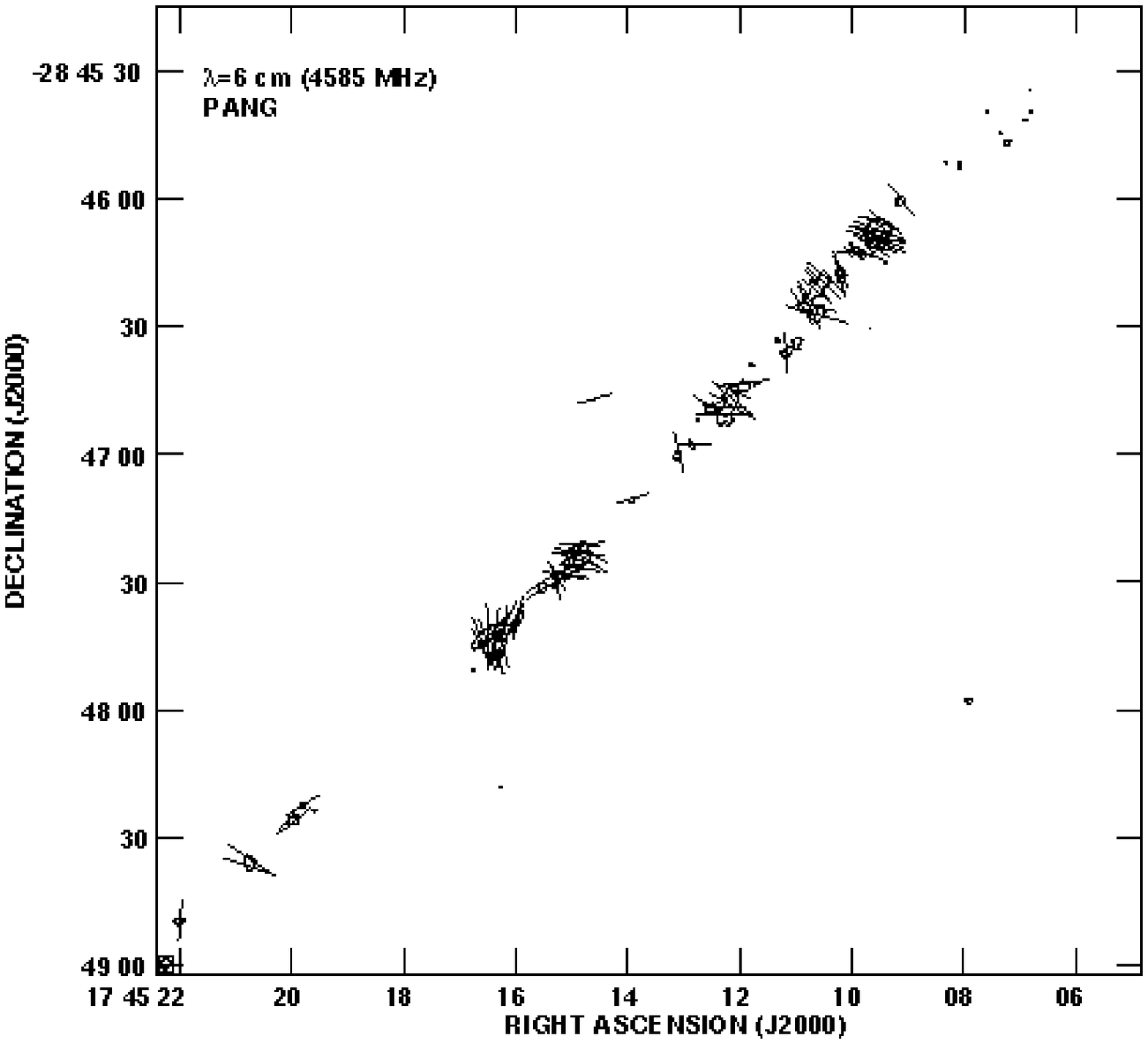}}
\begin{small}
\figcaption{\small Observed polarization angle of the \NT~at \l6 cm. The
contours represent polarized intensity, and the line segments are
scaled in length to reflect the magnitude of the polarized intensity
and show the orientation of the polarization angle of the detected
electric vector ($\theta$=0\arcdeg~is vertical and increases counterclockwise).}
\end{small}
\end{center}}

\vbox{%
\begin{center}
\leavevmode
\hbox{%
\epsfxsize=7.5cm
\epsffile{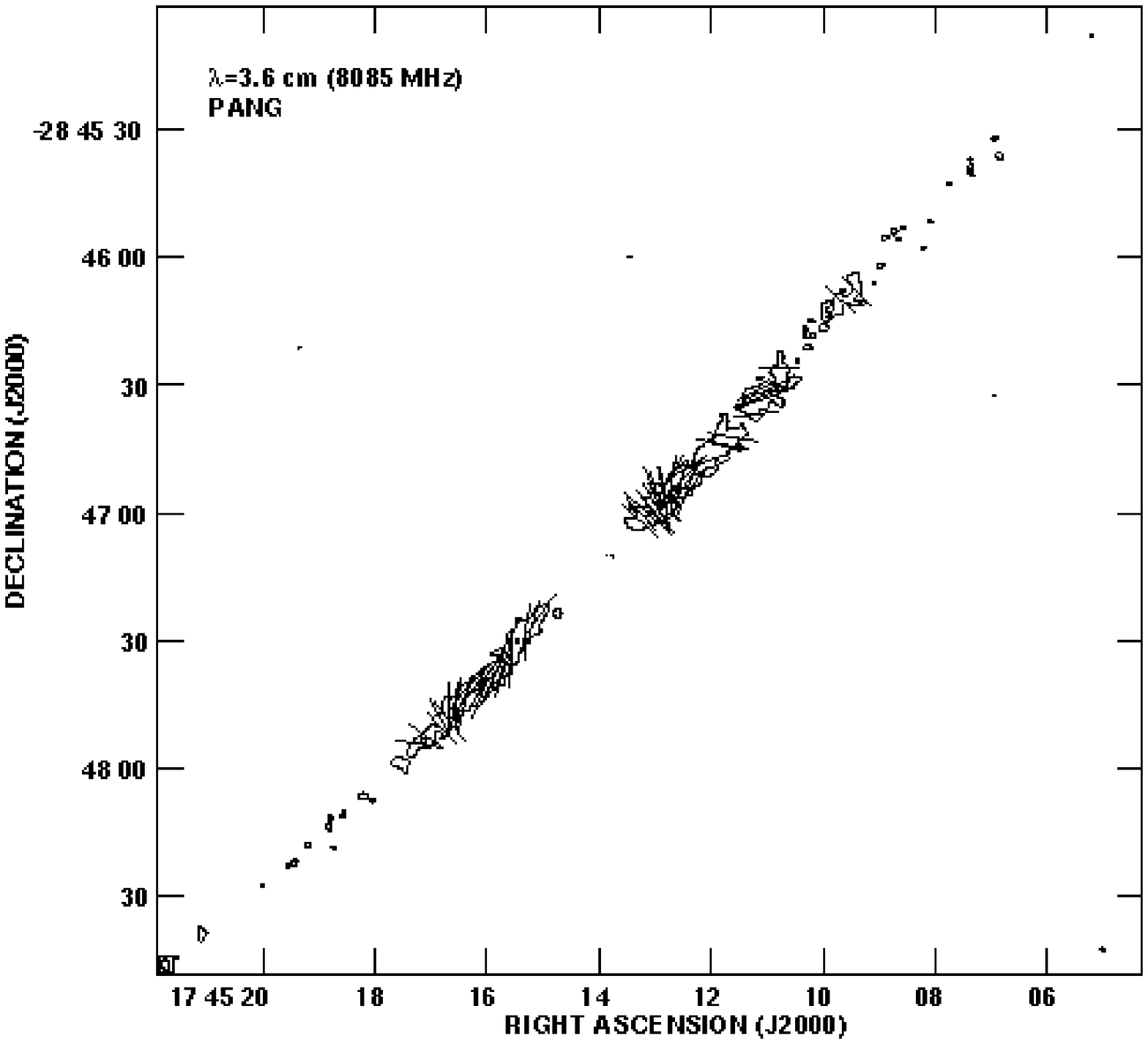}}
\begin{small}
\figcaption{\small Observed polarization angle of the \NT~at \l3.6 cm, with
contours and vectors in the same layout shown in Figure 12.}
\end{small}
\end{center}}

\vbox{%
\begin{center}
\leavevmode
\hbox{%
\epsfxsize=7.5cm
\epsffile{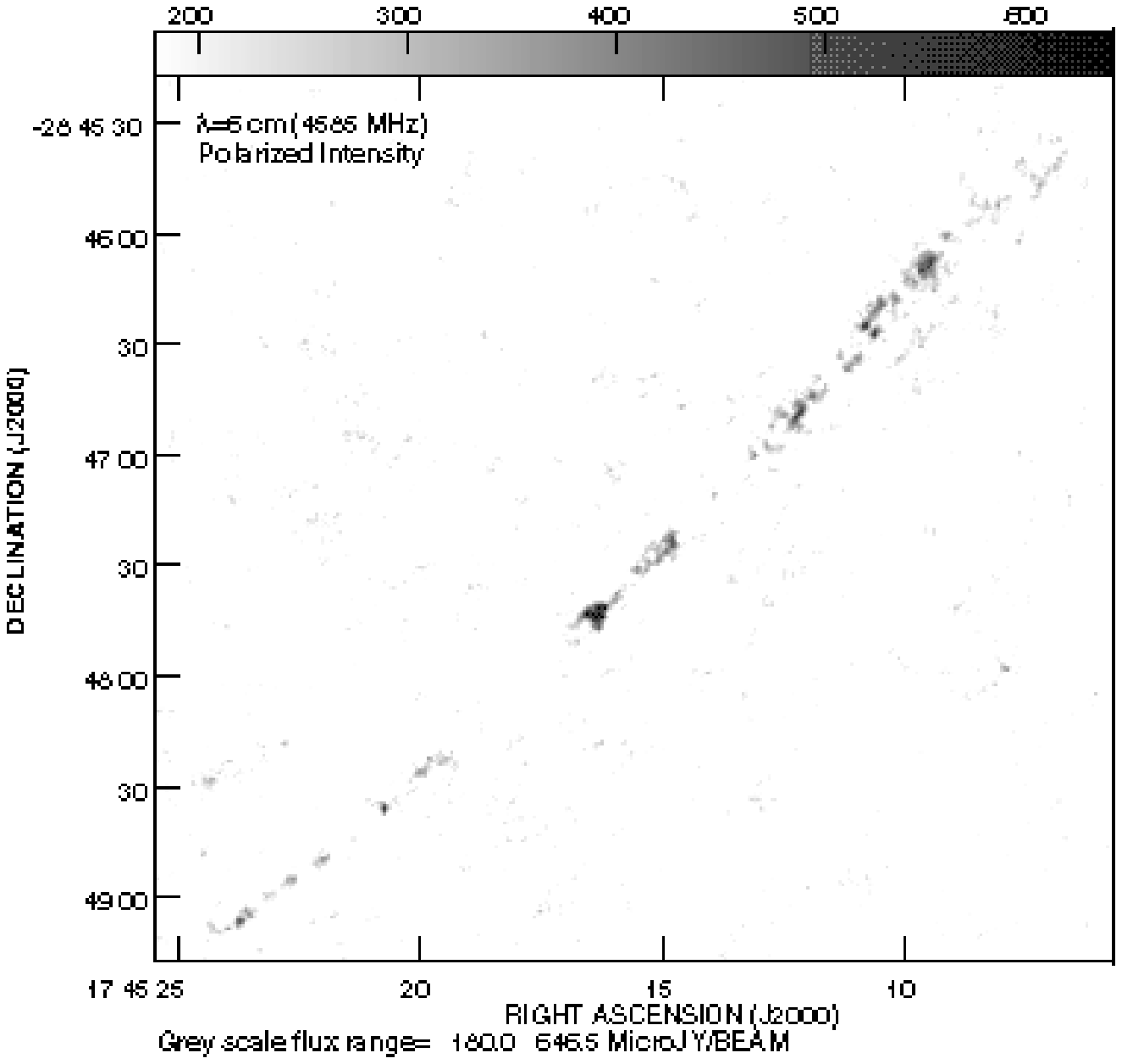}}
\begin{small}
\figcaption{\small Polarized intensity at \l6 cm of the Northern Thread, shown
in negative greyscale at a resolution of 2\farcs5 $\times$ 2\farcs0,
PA=0\fdg5. The image has been blanked at the level of 200 $\mu$Jy/beam. }
\end{small}
\end{center}}

\vbox{%
\begin{center}
\leavevmode
\hbox{%
\epsfxsize=7.5cm
\epsffile{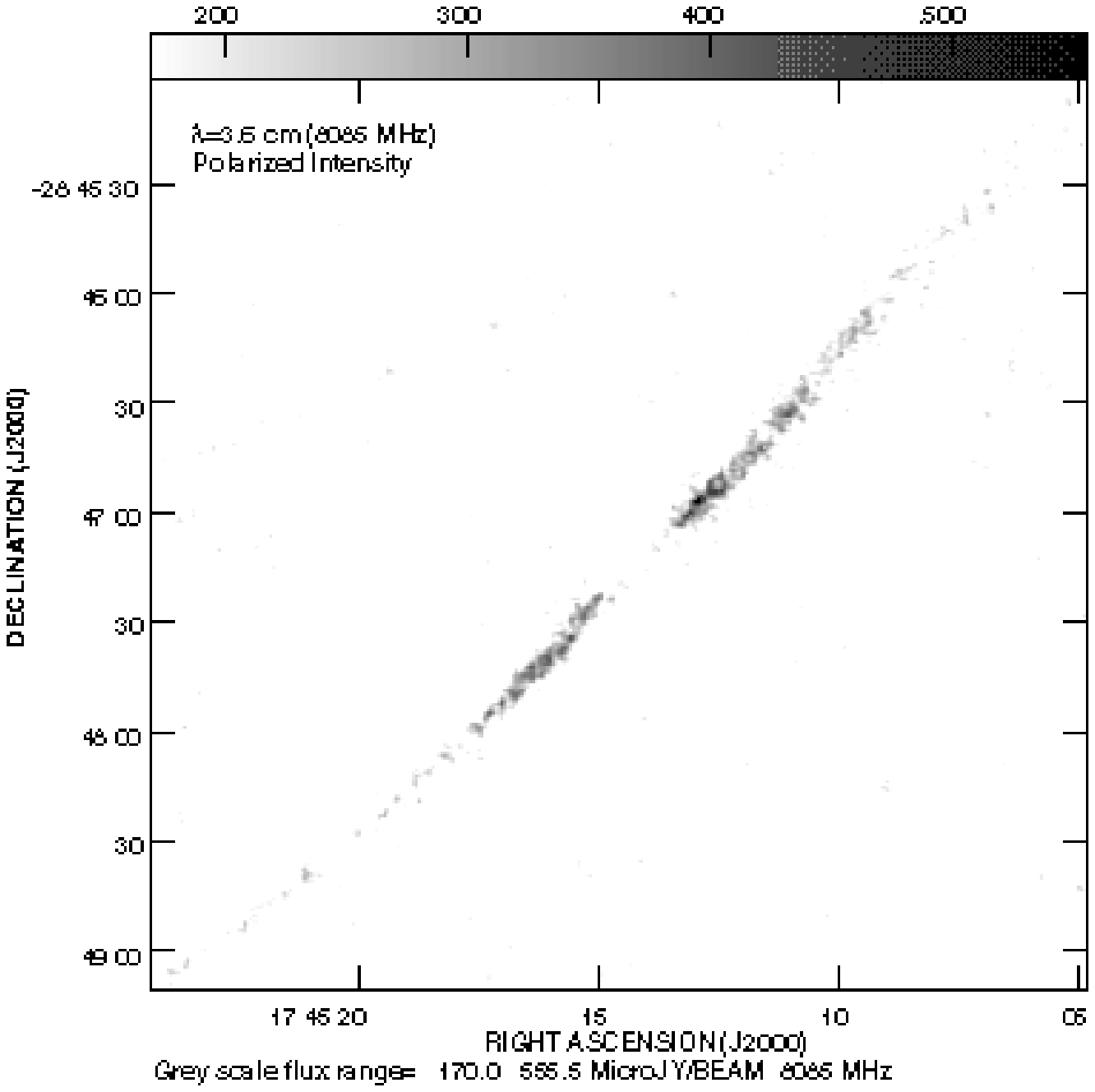}}
\begin{small}
\figcaption{\small Polarized intensity at \l3.6 cm of the Northern Thread shown in negative greysc
ale at a resolution of 2\farcs3 $\times$ 1\farcs9, PA=0\fdg8. The image has been blanked
 at the level of 170 $\mu$Jy/beam.}
\end{small}
\end{center}}

\subsubsection{Distribution of Polarization}

Figures 14 and 15 show the distribution of polarized intensity at \l6 cm
and \l3.6 cm, respectively.  In both cases, the polarized
emission extends along the ridge of total intensity, but unlike the total intensity,
its structure is patchy, and exhibits many gaps.  At both \l6 and \l 3.6 cm, polarized intensity is organized into
clumps along the filaments, at a scale of \ab5$-$10\arcsec~(0.2$-$0.4 pc). The peaks of
polarized intensity, typically \ab0.5 mJy beam$^{-1}$,
define a very narrow ridge along the filament.
The width of the distribution of polarized intensity appears to follow the pattern of
total intensity; it becomes very narrow to the E and abruptly
ends at (\a, $\delta$)$_{J2000}$=17 45 25,$-$28 49 10, which corresponds to the position
of the westernmost thermal Arched filament, intrinsically
unpolarized in radio emission. 
A splitting into two parallel filaments of the \l6 cm polarized
intensity at the W end of the \NT~is apparent in Figure 14. A similar
splitting of the Northern Thread is detected in total \l20 cm
intensity (Figure 2), however the positions of these splits do not
exactly correspond. The polarized \l6 cm split occurs between \a$_{J2000}$=17 45 00 to
17 45 10, $\delta$=$-$28 46 30, whereas at \l20 cm the splitting is located at \a$_{J2000}$=17 45 10 to
17 45 12, $\delta$=$-$28 45 30. Curiously, this sub-structure is not detected in polarized
emission at \l3.6 cm.  The polarized emission at \l3.6 cm (Figure
15) is more coherent than that at \l6 cm and does not
have such a patchy distribution. Similar patchy polarization
structure has been observed in the Snake
(Gray et al. 1995) and also in G359.54+0.18 (\yz~et al. 1997), and appears to be a common property of NTF's.

\subsubsection{Fractional Polarization}
The apparent fractional polarization as a function of length along the Northern
Thread for both \l6 cm and \l3.6 cm is shown in Figure 16. 
The upper limit of fractional polarization for radio synchrotron
emission under ideal conditions is 70\% for $\alpha$=$-$0.5 (Moffet 1975).  Typical
fractional polarization measures are determined to be in the range of
30$-$70\% for the other NTF systems observed (\yz~\& Morris 1988; Gray et al. 1995;
Yusef-Zadeh et al. 1997).  Here the maximum fractional polarization at \l3.6 cm is
probably artificially high ($>$90\%), owing to the fact that the largest scale structures of
the total intensity are not fully sampled by the shortest
spacings of the VLA, and therefore, a portion of the total flux
density is missing. At \l3.6 cm, the largest angular scale to which the VLA
is sensitive in the DnC array is on the order of 100\arcsec, whereas
the Northern Thread is \ab10\arcmin~in length. However, a qualitative trend is indicated in
Figure 16: the fractional polarization at \l3.6 cm appears to increase
from E to W along the Northern Thread. A more reliable calculation of fractional
polarization could be obtained by combining single dish and interferometer total intensity data. 

\vbox{%
\begin{center}
\leavevmode
\hbox{%
\epsfxsize=7.5cm
\epsffile{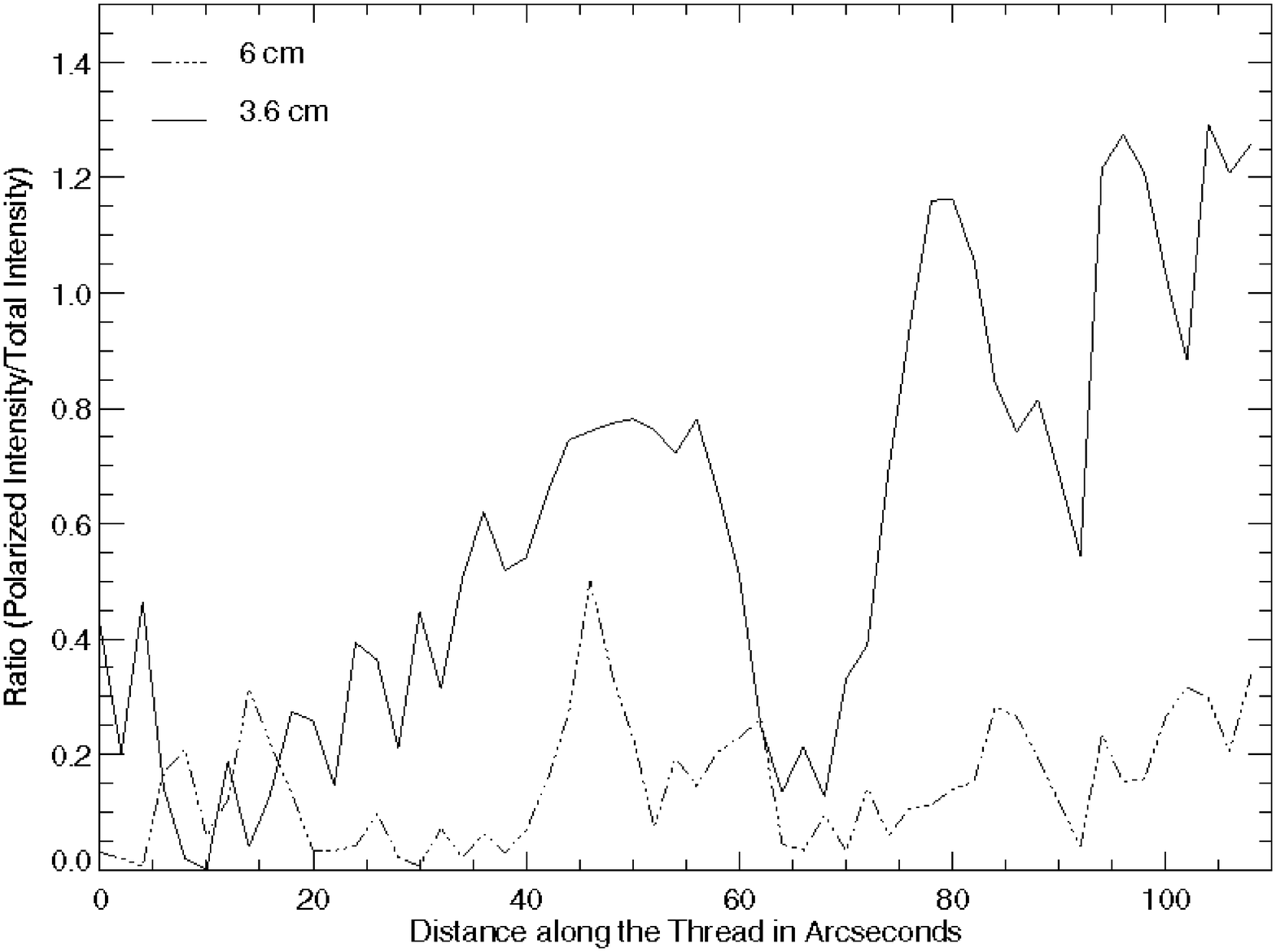}}
\begin{small}
\figcaption{\small Fractional polarization at \l3.6 and \l6 cm as a
function of length along the Northern Thread.}
\end{small}
\end{center}}

\subsubsection{Rotation Measure}  
Using the four observed frequencies at \l6 cm (4.585 \& 4.885 GHz) and
at \l3.6 cm (8.085 \& 8.465 GHz), the rotation angle as
a function of wavelength was fitted to a \l$^{2}$ law with the AIPS
algorithm RM. The resultant distribution of rotation measure (RM) in the \NT~is shown in false-color in Figure 17.  Figure 18 shows a sample fit of the polarization
angle versus $\lambda$$^{2}$ as a function of position for a small
region of the Northern Thread, demonstrating that no ambiguities in the
RM are present. 
The values for the RM toward the \NT~vary along the
source from 100 rad m$^{-2}$ to 2300 rad m$^{-2}$. Toward the center
of the Northern Thread, the RM values are in the range of 1000$-$2000
rad m$^{-2}$.  
Along the NW extreme of the \NT, where the total intensity
becomes wider and more diffuse, the RMs have lower values, 100$-$500 rad m$^{-2}$. 
\vbox{%
\begin{center}
\leavevmode
\hbox{%
\epsfxsize=7.5cm
\epsffile{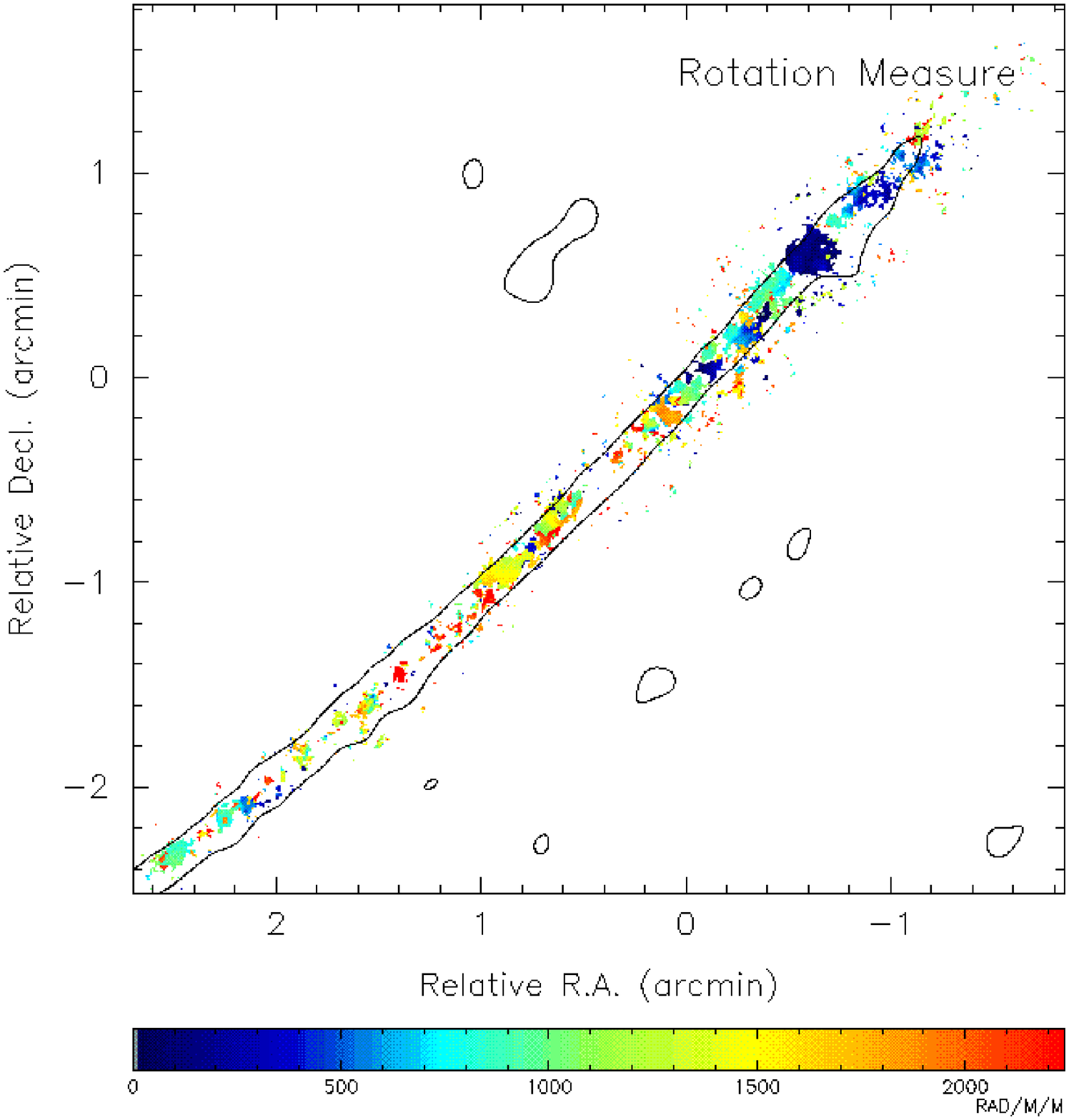}}
\begin{small}
\figcaption{\small Distribution of rotation measure (RM) along the Northern Thread in
false color scale with units of radians m$^{-2}$, with a resolution of
2\farcs5 $\times$ 2\farcs0. The superposed contour represents the 3
mJy beam$^{-1}$ level of the \l6 cm total intensity image of the
Northern Thread, which has been smoothed to a resolution of 6\farcs2
$\times$ 5\farcs5. The RM has been blanked for errors in the polarization angle that are great
er than 10\%.}
\end{small}
\end{center}}

\vbox{%
\begin{center}
\leavevmode
\hbox{%
\epsfxsize=7.5cm
\epsffile{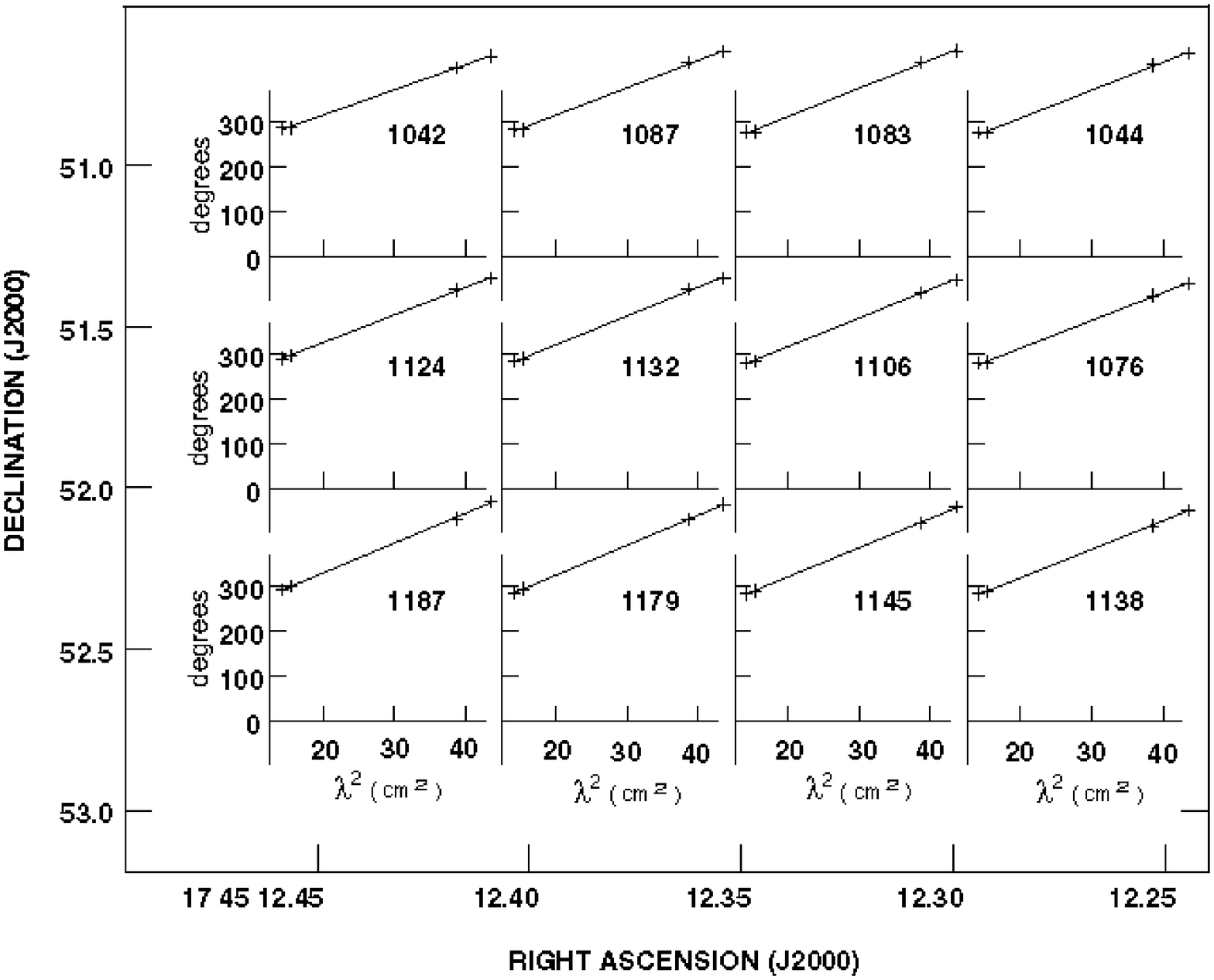}}
\begin{small}
\figcaption{\small Fit to the polarization angle versus wavelength squared are
shown as a function of position for a small region in the NW portion of the \NT, where the rot
ation measures are low. The fits are plotted for every third pixel.}
\end{small}
\end{center}}
\vbox{%
\begin{center}
\leavevmode
\hbox{%
\epsfxsize=7.5cm
\epsffile{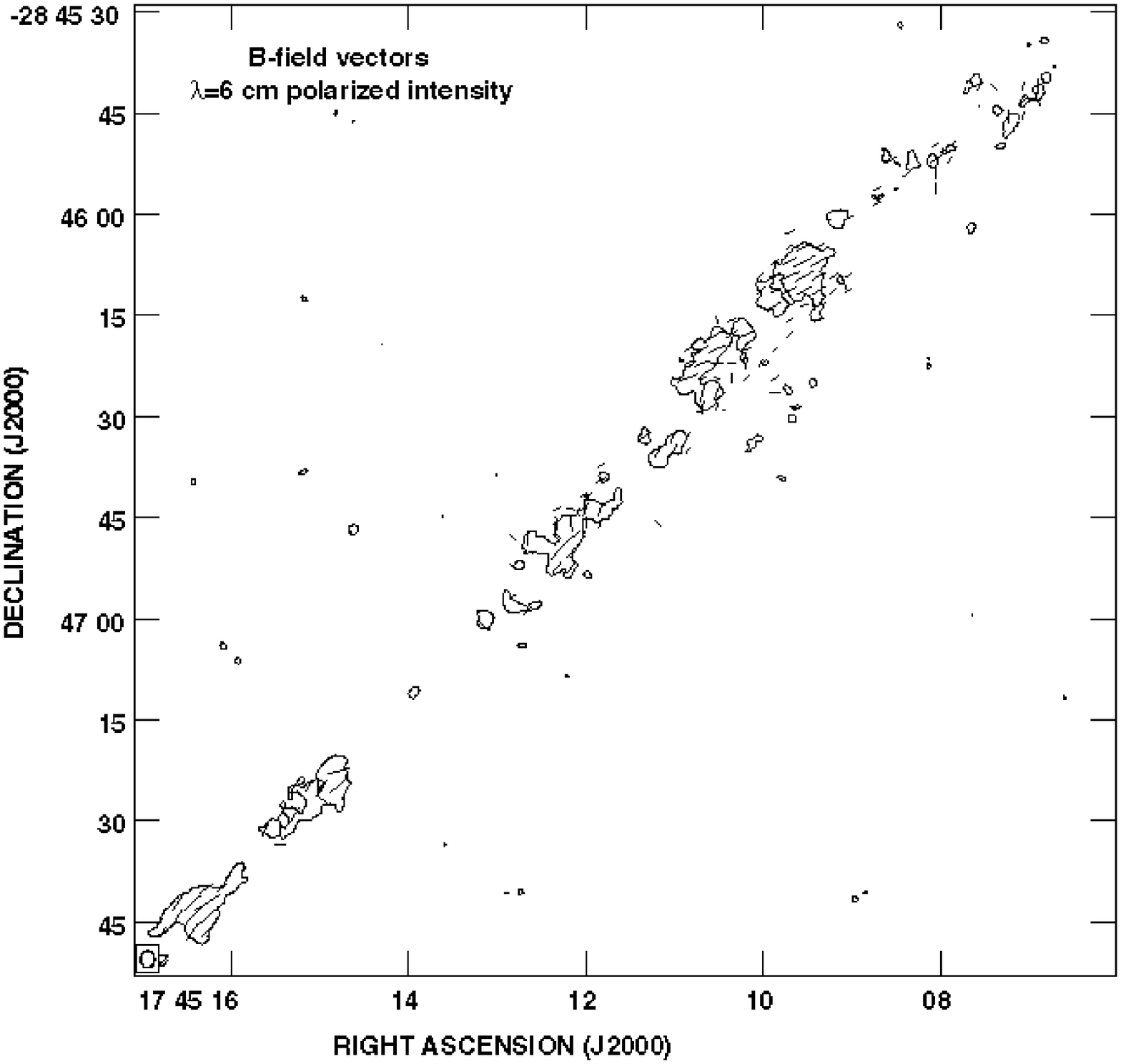}}
\begin{small}
\figcaption{\small Vectors representing the intrinsic orientation of the
magnetic field in the Northern Thread are shown superposed on the
contours of polarized intensity at \l6 cm.}
\end{small}
\end{center}}

\subsubsection{Intrinsic Magnetic Field}
The vectors in Figure 19 represent the orientation of the magnetic
field intrinsic to the \NT~obtained by correcting the observed
position angles for Faraday rotation. 
The magnetic field is predominantly aligned parallel to the \NT,
as expected from the studies of other NTF's.  In
a few positions along the filament, there are deviations in the magnetic
field orientation, up to 45\arcdeg~from alignment with the orientation
of the NTF. \yz~et al. (1997) observe a similar effect in
G359.54+0.18.  They point out that such positions correspond to the
edges of the ``clumps'' of polarization, where the
RM gradient
is also very high. In such places in the Northern Thread, the inferred direction of the
magnetic field is not as reliable because the emission is nearly
depolarized and the S/N is thus degraded.

\subsection{Southern Thread}
Although the Southern Thread is not detected in total intensity
at this frequency, a low level of polarized intensity is
detected. 
Figure 20 shows the distribution of polarization at \l3.6 cm of the
Southern Thread. Using the upper limit for the total intensity at
\l3.6 cm,
we derive a limit for the fractional polarization in the Southern
Thread of $\gtrsim$30\%. The splitting
of the Southern Thread into two filaments is also apparent in
the polarized intensity. For this filament, we did not attempt to derive
the intrinsic magnetic field orientation
since polarized emission is only detectable at two of the four observed
frequencies.  The polarization angle of the \l3.6 cm emission is shown in
Figure 21.   
\vbox{%
\begin{center}
\leavevmode
\hbox{%
\epsfxsize=7.5cm
\epsffile{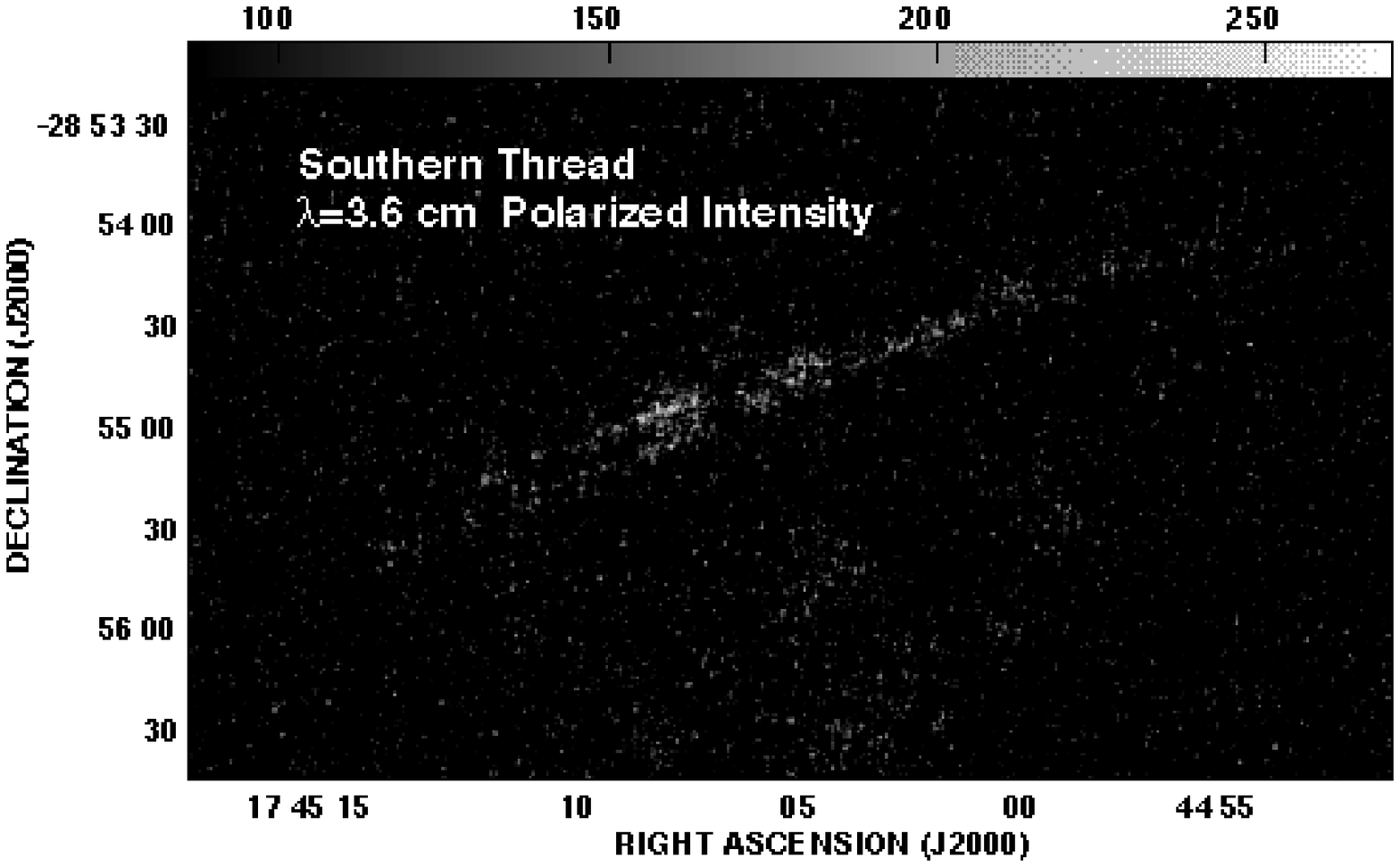}}
\begin{small}
\figcaption{\small Polarized intensity at \l3.6 cm arising from the Southern
Thread shown in greyscale, at a resolution of 2\farcs0 $\times$ 1\farcs5. }
\end{small}
\end{center}}

\vbox{%
\begin{center}
\leavevmode
\hbox{%
\epsfxsize=7.5cm
\epsffile{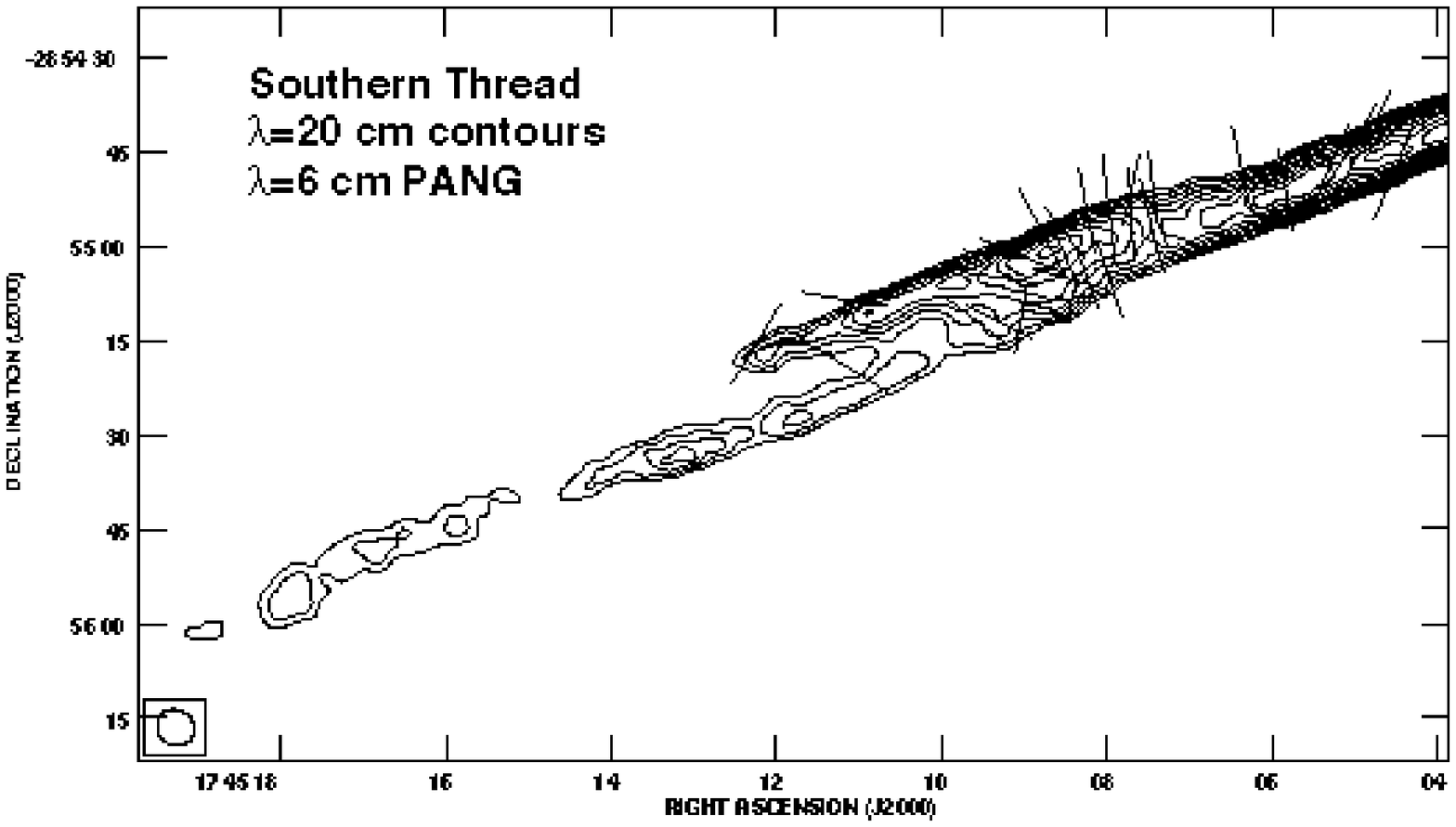}}
\begin{small}
\figcaption{\small Observed polarization angle of the Southern Thread at \l3.6 cm, with vector orientation as in Figure 12.}
\end{small}
\end{center}}

\section{DISCUSSION OF TOTAL INTENSITY RESULTS}
\subsection{Synchrotron Parameters}
Using the formalism of Moffet (1975), we can derive an equipartition
magnetic field strength from the radio continuum luminosity in the
Northern Thread.  Assuming that the filament is a cylinder,
with length 700\arcsec~(30 pc) and radius 6\arcsec~(0.25 pc), we
calculate an
equipartition magnetic field strength of B$_{eq}$\ab140 $\mu$G, comparable to the values derived for
the NTF's in the Radio Arc using equipartition methods (Tsuboi et al. 1986; Yusef-Zadeh et
al. 1987a; Reich 1994). The break in 
the spectrum at $\nu_{cutoff}$=4.585 GHz (\l6 cm) in Figure 10 implies a typical 
electron energy of \ab1 GeV,  and a synchrotron lifetime of the electrons 
in the Northern Thread, $\tau_{syc}$\ab4 $\times$10$^5$ years. 

However, the
magnetic field is unlikely to be in equipartition with the
relativistic particles. If the particles are in
equipartition with the magnetic field, then 
these filaments, which presumably trace the field, will be strongly distorted by interactions with the
tumultuous medium in which they are embedded.  The NTF's are
observed to extend
for up to 40 pc with no signs of local disturbances. 
Therefore, the highly organized field probably
strongly dominates the energy density on the observed scales, in which
case the magnetic field is substantially larger than the equipartition 
value, and the synchrotron lifetime correspondingly shorter.
A more accurate estimate of the magnetic field strength in the
Northern Thread may be obtained by equating the ram pressure or
turbulent pressure of the molecular cloud that intersects the Northern 
Thread with the magnetic pressure of the NTF's (Morris \& Yusef-Zadeh 1989).
This argument rests on the fact that, in most cases, the NTF's appear
rigid in their orientations 
at places where they are apparently interacting with the turbulent
molecular gas. We deduce a lower limit of \ab1 mG for the Northern and Southern
Threads has been estimated using this method (Morris \& Yusef-Zadeh
1989), and similar arguments yielding similar field strengths have been applied to other NTF's.  

\subsection{Spectral Index Variation along the Northern Thread} 
According to the model of Serabyn \& Morris (1994), the electrons
which stream along the NTF in the Radio Arc are injected into the NTF's
at the site of reconnection, where the ionized gas, molecular cloud and NTF's
intersect. In the case of the Northern Thread complex, similar elements are
present: (1) the magnetic Northern Thread, (2) the ionized Arched Filaments
(Yusef-Zadeh 1986), which lie on the surface of (3) a molecular cloud 
with peculiar velocity of $-$30 \kms~(Serabyn \& \gusten~1987).  The site 
of acceleration would therefore be expected to lie at the intersection of
the Arched Filaments and the Northern Thread.  Thus, a steepening of the
spectral index westward along the Northern Thread from that point should
be observed due to electron energy losses.  However, as discussed in
$\S$2.2.1, no significant steepening of the spectral 
index westward along the length of the Northern Thread is detected.
 
A possible explanation for the lack of steepening is that the electrons may be diffusing along the
filament more rapidly than they are losing energy via synchrotron
radiation. To calculate the electron diffusion timescale, we
assume that the Alfv\'{e}n speed provides an upper limit on the speed at
which the electrons diffuse along the filament (Alfv\'{e}n 1942).
The diffusion timescale has the following dependences on the electron
density, n$_e$, and the magnetic field, B: $\tau_{diff}$ $\propto$
$\frac{[n_e(cm^{-3})]^{1/2}}{B}$. For n$_e$ $\lesssim$
1 cm$^{-3}$, and for a filament length of 30 pc, $\tau_{diff}$$\lesssim$1.5
$\times$10$^{4}$ years ($\case{1~mG}{B_{eq}}$). On the other hand,
the synchrotron lifetime, $\tau_{syc}$=1.7 $\times$ 10$^4$ years ($\case{1~mG}{B}$)$^{2}$.
Since the timescales for synchrotron losses and diffusion are
comparable, it is difficult to draw conclusions
about the location of acceleration or reacceleration of the electrons along the
filament, as a variation in the spectral index would be small,
and not detectable within the current errors.  
A lack of variation of the spectral index is also observed
along the Sgr C NTF (Kassim et al. 1999), and 
the Snake (Gray et al. 1995). The fact that the electrons are 
streaming along the NTF at a similar rate as they radiate their
energy via synchrotron, raises the question: why do the NTF's terminate
abruptly, fading quickly into the diffuse background?
 
\subsection{Divergence of the Magnetic Field}
One possible answer is that the divergence of the magnetic field,
evidenced by the increase in width of the Northern Thread along its 
length, causes a reduction in the synchrotron emissivity. The observed 
intensity of synchrotron emission as a function of filament width, I(w), 
can be modelled as arising from a cylindrical magnetic flux tube of width
w; this relation can be derived as follows. I(w) is proportional to the 
product of the synchrotron emissivity per particle, $\epsilon$, the number 
density of relativistic particles in the cylinder, n$_e$, and the
depth of the cylinder, assumed equal to the width, w: I(w) $\propto$ 
$\epsilon$~$\cdot$~n$_e$~$\cdot$~w.  Further, $\epsilon$ $\propto$ B$^{3/2}$,
where B is the magnetic flux density, and in turn, if the total magnetic
flux within a filament is conserved, B $\propto$ w$^{-2}$.  Also,
if the relativistic particle flux along an NTF is constant, n$_e~\propto$ 
w$^{-2}$.  
\vbox{%
\begin{center}
\leavevmode
\hbox{%
\epsfxsize=7.5cm
\epsffile{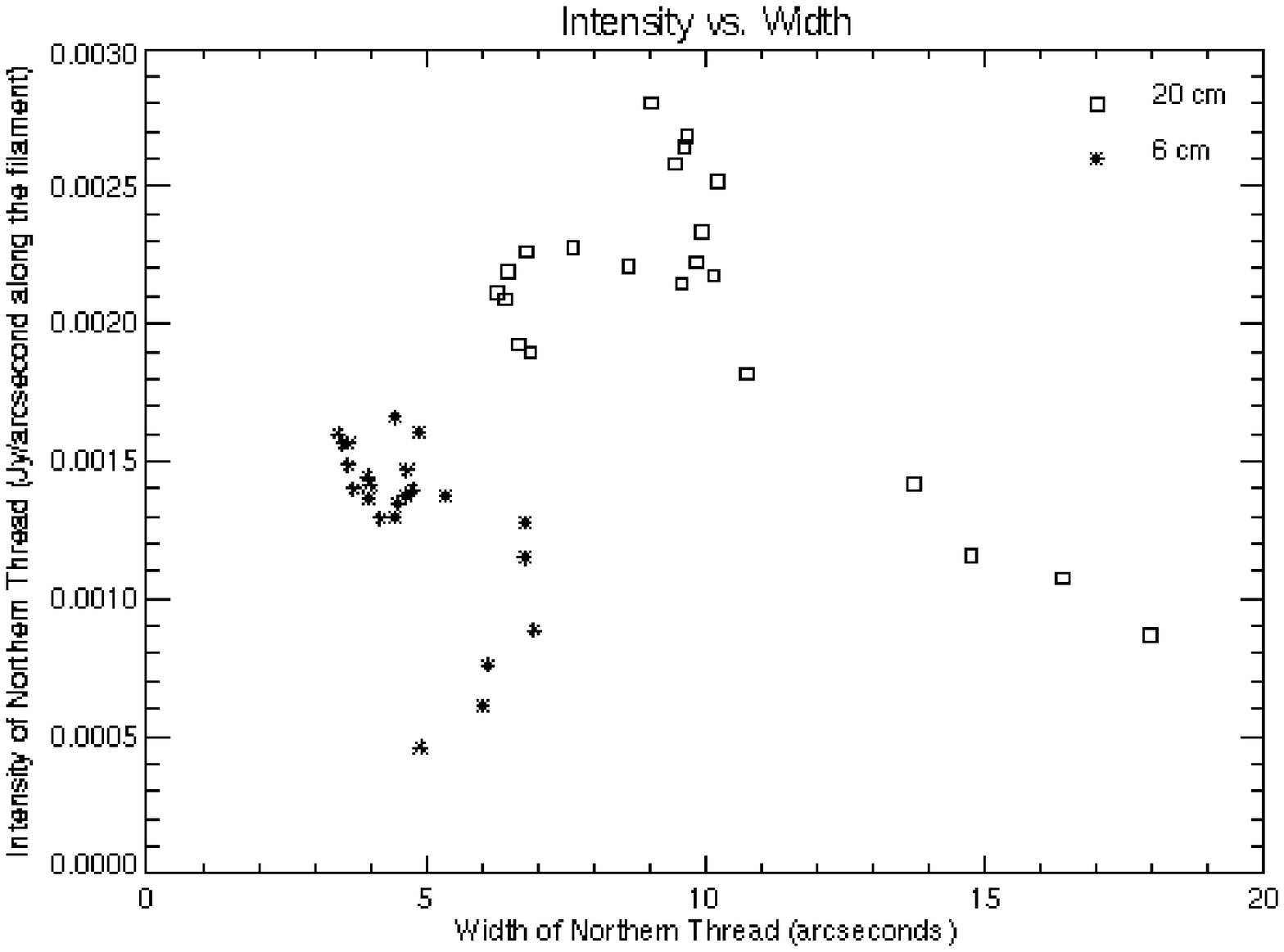}}
\begin{small}
\figcaption{\small Plot showing intensity versus width of the Northern
Thread for \l6 cm (stars) and \l20 cm (squares). }
\end{small}
\end{center}}

These relations imply that I(w) $\propto$ 
w$^{-4}$. To assess the applicability of this scaling, we plot the \l20
cm and \l6 cm intensity of the Northern Thread as a function of its width
in Figure 22. Although the data indicate that the intensity decreases 
with increasing width, more convincingly at \l20 cm than at \l6 cm, the slope 
of the decrease is not as steep as the predicted w$^{-4}$. 
One way to account for the weaker dependence is to note that the cross-section 
of a given flux tube may be elliptical rather than circular, and that 
the apparent change of width of the filament could correspond to variations 
in the orientation of the major axis of that ellipse (i.e., similar
to a thick,
twisting ribbon), rather than to a compression or divergence in the
magnetic field.  If the variation of intensity 
with width is due only to the change of orientation of the cross-section, 
then the magnetic field does not depend on the width, and the surface brightness depends 
only on the column density of relativistic particles, which goes roughly
as w$^{-1}$

Yet another possibility to explain the abrupt fading of the NTF's is that the decline of the ambient particle density with height above the
plane, coupled with fewer sources of turbulence, (thus fewer density
inhomogeneities), leads to a dramatic increase in drift velocity, 
and thus to the reduction of the density of relativistic synchrotron-emitting 
electrons. 
 
\subsection{Interaction with the Arched Filaments}

The physical nature of
the interaction between the Northern Thread, the Arched Filament H II regions, and the
underlying molecular cloud in this region is uncertain. Marked
discontinuities in brightness of the Arched Filaments are apparent in
Figure 23 where the Northern Thread intersects them. Figures 24 a \& b
show slices parallel to the Arched Filaments W1 and W2, over a distance
of \ab40\arcsec~from N to S, roughly centered on the position where
the Northern Thread intersects them. As evident in Figure 24a, the
portion of W1 lying to the S of where the Northern Thread intersects
it is dramatically brighter than the portion to the N of the NTF.  In Figure 24b, there is a
marked decrease in brightness of W2 S of the position where the
Northern Thread intersects the thermal gas.

\vbox{%
\begin{center}
\leavevmode
\hbox{%
\epsfxsize=7.5cm
\epsffile{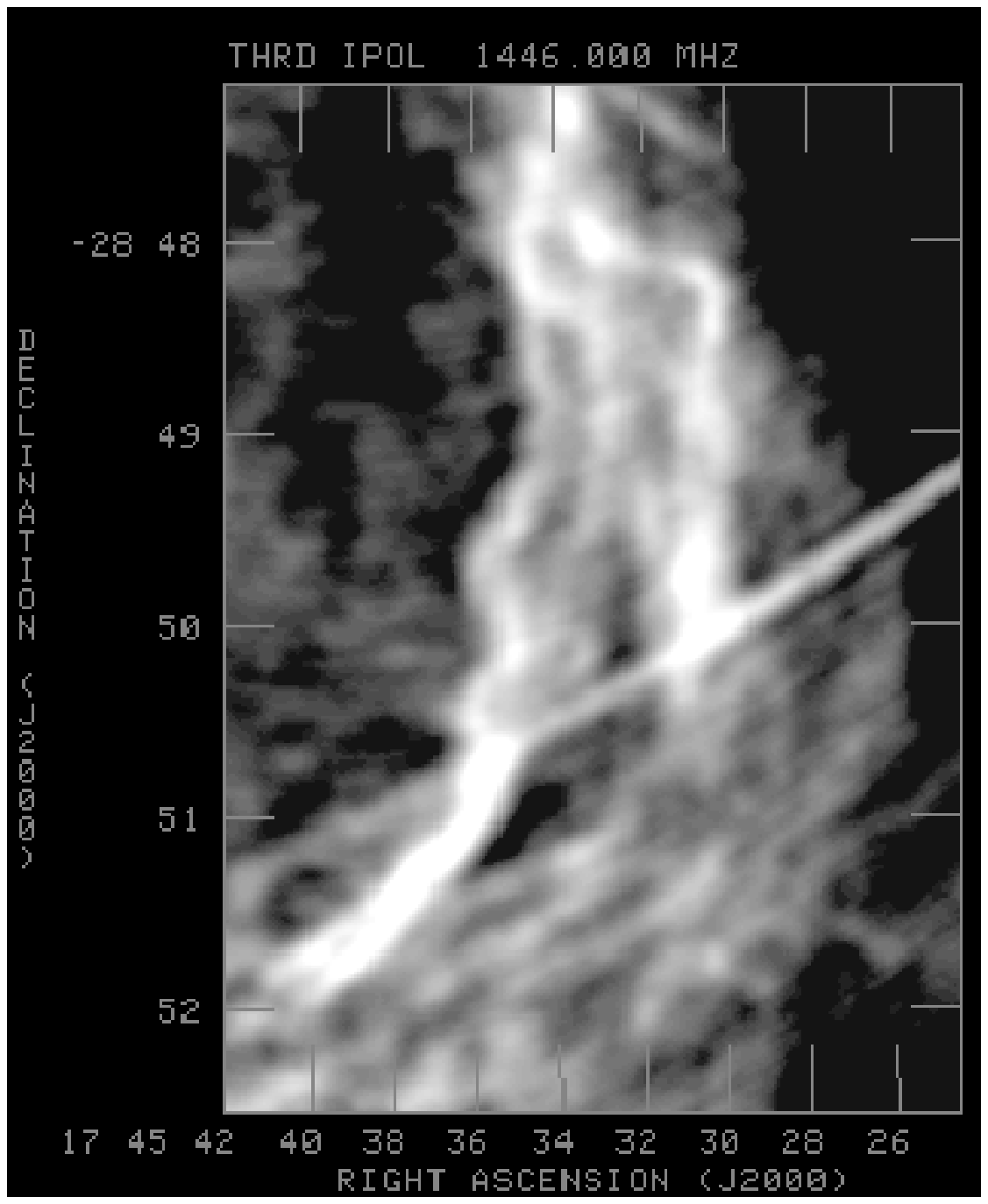}}
\begin{small}
\figcaption{\small A detail from \l20 cm image in Figure 2, highlighting
the positions where the Northern Thread intersects the Arched
Filaments W1 and W2.}
\end{small}
\end{center}}

There is evidence in other NTF/ionized gas systems that the
observed changes in brightness of intersecting sources imply a
physical interaction between the sources.  However, in the case of
the Radio Arc NTF's, it is the brightness and continuity of the NTF's
which change significantly at intersections with the Sickle H II
region, rather than the thermal structures (Yusef-Zadeh \& Morris
1987c). At this interface, magnetic reconnection has been suggested as
the mechanism for acceleration of the electrons along the NTF's of the Radio Arc (Serabyn \&
Morris 1994), but similar evidence for this mechanism in the case of
the Arched Filaments and Northern Thread is not present.

\vbox{%
\begin{center}
\leavevmode
\hbox{%
\epsfxsize=7.5cm
\epsffile{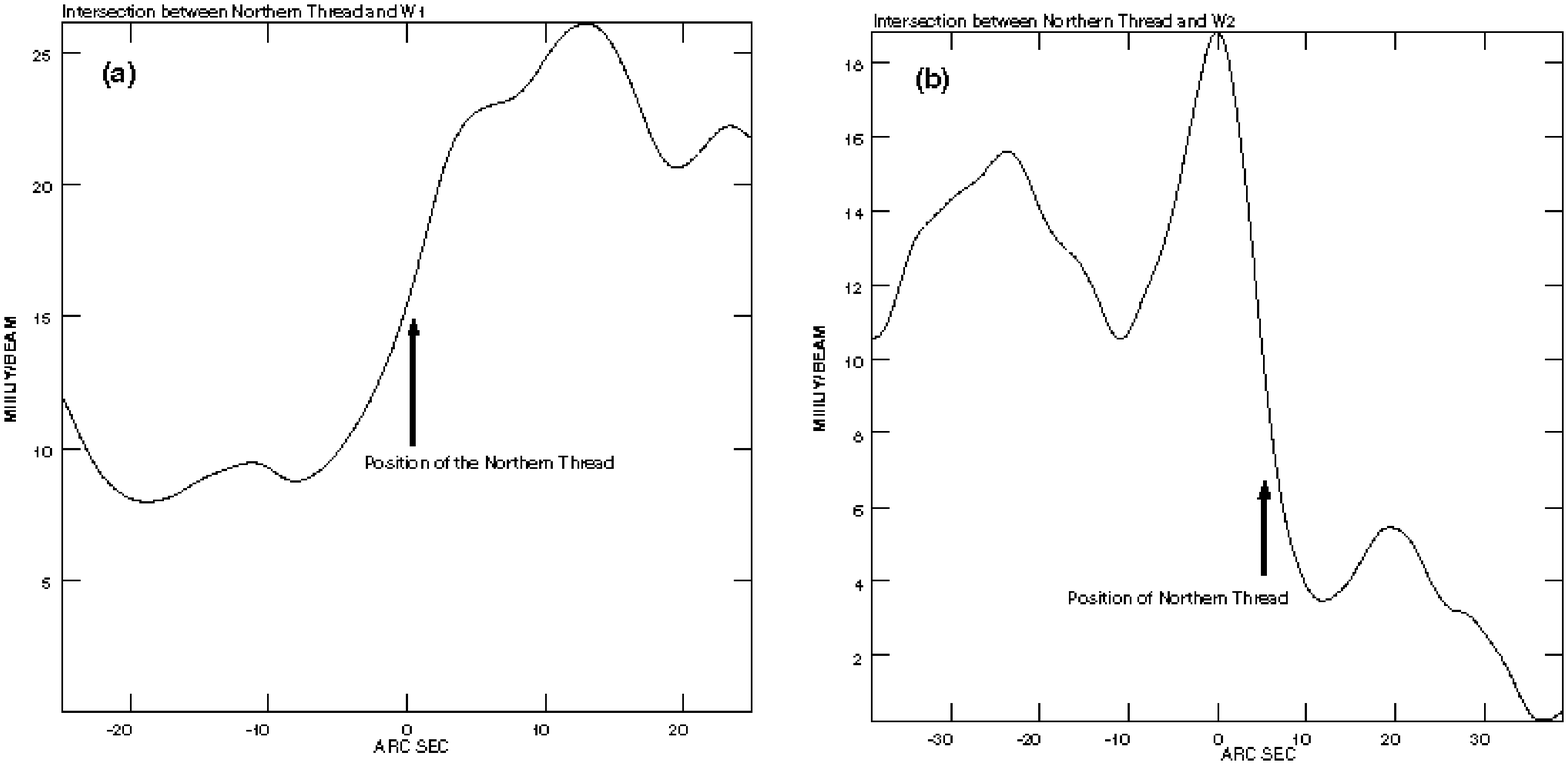}}
\begin{small}
\figcaption{\small Slices in \l20 cm total intensity parallel to the Arched
Filaments (a) W1 and (b) W2, over a distance of \ab40\arcsec~from N to
S, at positions where the Northern Thread intersects the thermal gas.}
\end{small}
\end{center}}

\subsection{Motion of the Threads}
The 3.9-year time base of the \l20 cm observations was used to put
constraints on the transverse motion of the Northern and Southern
Threads. The cosmic string model of Chudnovsky et al. (1986) predicts
that NTF's should have relativistic, transverse motion. In order to detect possible transverse motion, we
implemented the cross-calibration method proposed by Masson (1986).  A difference image,
made from differencing the data at two epochs,  showed no
evidence for transverse motion of the Northern and Southern
Threads. We calculate upper limits of 0.03c and 0.06c for the lateral motion of any portion
of the Northern and Southern Threads, respectively, thereby
constraining relativistic lateral motion to the unlikely case of being
entirely in or near the plane of the sky.

\section{DISCUSSION OF POLARIMETRY RESULTS}
\subsection{Sources of Depolarization}

The observed polarization is extremely patchy, but similar small gaps
in the polarization structure is observed in the other NTF's.  In the
case of the Radio Arc, the distribution of the polarized
intensity has been attributed to depolarization by small-scale thermal
media observed as large scale ``helical'' emission features apparently
surrounding the Radio Arc (Yusef-Zadeh et al. 1984; Inoue et
al. 1989). No thermal 
structures which could be causing the patchy depolarization are detected
toward the \NT, other than the Arched Filaments. The
abrupt end of the polarized emission at the W extent of the Northern
Thread can be attributed to the likelihood of the Arched Filaments
lying at least partially front of the Northern Thread. 
 
As apparent from Figures 14 and 15, the polarized emission at \l3.6 cm
is more coherent and not as patchy as the \l6 cm polarized emission.  
This is consistent with a \l$^2$-dependence on rotation angle, which
causes the beam depolarization to be larger at longer wavelengths. In addition, the overall high
fractional polarizations and the large rotation of the plane of
polarization (100$-$300\arcdeg; see Figure 18) indicate that the Faraday
rotation is predominantly external to the filament.     
Internal Faraday rotation cannot rotate the plane of polarization by
more than 90\arcdeg~without substantial depolarization (Burn 1966). 

\subsection{Faraday Rotation}
In order
to try to locate the Faraday rotating medium by associating it with either
the source or with structures in the intervening medium, such as those
detected by Inoue et al. (1989), we looked for coherent structures in the
distribution of the rotation measure. None were detected. However, as described above, the large rotations
are probably not due to internal Faraday rotation.  Therefore, the Faraday screen
is located near the distance of the NTF's (i.e., at the Galactic center)
or in a region somewhere along the line of sight.  
The highest values of RM observed in the \NT~(1000$-$2500 rad m$^{-2}$)
are similar to what has been observed toward other NTF's. Gray et al. (1995) find RMs of 2000$-$5000 rad m$^{-2}$
along the Snake, with values increasing toward the Galactic plane. The
RMs toward the linear filaments in the Radio Arc, located much closer
to the plane than the \NT, are 2000$-$5500 rad m$^{-2}$ (\yz~\&~Morris 1987b).
Yusef-Zadeh et al. (1997) argue that the high RMs they observe in
G359.54+0.18 (up to 5000 rad m$^{-2}$) are more than two
orders of magnitude larger than for sources outside of the inner
Galaxy, and that the Faraday rotations are likely occurring closer to the Galactic center.  

\section{Ubiquity of the Magnetic Field in the Galactic Center}
As discussed in Section $\S$2.1.2, a number
of linear structures similar to the known NTF's are apparent in
Figure 2.  These ``streaks'' are aligned
essentially perpendicular to the Galactic plane,
but extend for only 5$-$7 pc, as opposed to the 10's of pc extent of the NTF's.  The considerably shorter length might indicate that
they are at a different stage in their development than the NTF's. We
note that the
multiplicity of NTF's and related linear features, such as streaks,
may far surpass what we can observe with current sensitivity. 
The relatively uniform orientation of all of these features
(perpendicular to the plane) is
consistent with a simple, polodial geometry for the magnetic field
within the central few hundred parsecs of the Galaxy (Morris 1994).    
In fact, the multiple parallel filaments in both 
the Northern and Southern Threads (see Figs. 2 \& 6) indicate that
the synchrotron illumination of the magnetic field structure 
shifts to adjacent flux tubes along the length of the NTF's, perhaps
as a result of the lateral motion of the source of relativistic
particles with respect to the ambient, pervasive field. 
Such a magnetic field can provide
the support via magnetic pressure to account for the narrowly confined
NTF's, since magnetic pressure acts perpendicular to the field
orientation. Models which
rely on localized field enhancements (i.e. a local force-free
field/current configuration) encounter difficulties providing a mechanism for keeping
the NTF's so highly collimated and with preventing their rapid expansion.  
In addition, the slight curvatures of the Northern Thread and the SgrC
NTF are in opposite directions. This raises the possibility of a
general divergence of the polodial field at the Galactic center with
increasing Galactic latitude.  Indeed, allowing for projection
effects, the large-scale curvature of all known NTF's is consistent
with such a divergence. 

\section{Conclusions}
A multi-frequency study of the Northern and Southern Threads was
carried out with the VLA using data
from a number of different epochs and configurations.  The following
conclusions have been made:

(1) A \l20 cm image is presented of a 60 pc region of the inner
    Galaxy, mostly at positive latitudes and longitudes.   The Sgr A West complex and the Northern
    and Southern Threads are prominent features. In addition, a number
    of new thermal and linear non-thermal sources are observed and
    discussed. 

(2) The Southern Thread shows remarkable substructure at \l20cm, but
    does not stand out against the diffuse halo of SgrA at \l3.6 cm
    and \l6 cm, except in \l3.6 cm polarized intensity.   

(3) For the first time, the \NT~is resolved. In the total \l6 cm
    intensity image its width varies along
    its length, from $\lesssim$4\arcsec~(0.16 pc) at its narrowest
    position to \ab12\arcsec~(0.5 pc) at its most broadened position.  

(4) The spectral index of the \NT~is relatively 
   steep (\a=$-$0.5 between \l90 and \l6 cm), with
   increasing steepness at the shorter wavelengths (\a=$-$2.0 between \l6
   and \l2 cm).  The lack of significant variation of the spectral
   index along the filament prevents us from locating the source of acceleration of the
   emitting electrons. 

(5) The polarized intensity arising from the \NT~is patchy and
    discontinuous along its length at \l6 cm, and is more coherent at
    \l3.6 cm. The fractional polarizations at \l3.6 cm are
    quite large in some places but likely to be upper limits because the total intensity on the largest scales may not be well sampled by the VLA.
     
(6) Values of the rotation measure range from 100$-$2300 rad m$^{-2}$ 
    along the \NT. The larger RM's are consistent with those observed toward
    NTF's. No coherent structures in
    the distribution of RM are observed which would give clues about
    the location of the Faraday rotating medium.  

(7) The intrinsic magnetic field in the \NT~is primarily aligned with
    the filament.  

(8) The Northern Thread shares all of the characteristics
    of the rest of the known NTF's: strong linear polarization, falling
    spectrum, large RM's in some locations, and an intrinsic magnetic field aligned along the
    filament length.  Although the data were less conclusive on the
    Southern Thread, it, too, appears to show many of these
    characteristics.  
The large number of roughly aligned filamentary features in the 
$\lambda$20 cm image evokes the presence of a large-scale, dipole field 
permeating the inner 150 pc of the Galaxy.

\acknowledgements{We would like to thank G. Taylor of NRAO in Socorro for assistance with the polarization data, and M. Goss at NRAO in
Socorro for assistance in imaging the \l20 cm data, and helpful comments on the manuscript.  CCL acknowledges the RA
Mentorship Fellowship at UCLA for support during the preparation of this manuscript.}

\end{document}